\documentclass[showpacs,aps,prd,reprint,nofootinbib]{revtex4-1}
\usepackage[colorlinks=true, pdfstartview=FitV,citecolor=blue, bookmarksnumbered=true, breaklinks]{hyperref}
\usepackage[dvipdfmx]{graphicx}
\usepackage{amsmath,amssymb,bm,color,longtable,amsfonts,slashed}

\usepackage{dcolumn}
\usepackage{bm}
\usepackage{color} 
\newcommand{\pmat}[1]{\begin{pmatrix} #1 \end{pmatrix}}

\date{\today}

\begin{document}
\title{Quark-diquark potential and diquark mass from Lattice QCD}

\author{Kai Watanabe}
\email[]{kaiw@rcnp.osaka-u.ac.jp}
\affiliation{Research Center for Nuclear Physics, Osaka University, Japan}

\newcommand{\Sect}[1]{Section~\ref{#1}}
\newcommand{\Fig}[1]{Fig.\ref{#1}}
\newcommand{\Table}[1]{Table~\ref{#1}}
\newcommand{\Eq}[1]{Eq.~(\ref{#1})}
\newcommand{\Appendix}[1]{Appendix \ref{#1}}
\newcommand{\rmII}{\MakeUppercase{\romannumeral 2\ }}
\newcommand{\rmIII}{\MakeUppercase{\romannumeral 3\ }}

\begin{abstract}
We propose a new application of lattice QCD  to calculate the quark-diquark potential, diquark mass and quark mass required for the diquark model.
As a concrete example, we consider the $\Lambda_c$ baryon and treat it as a charm-diquark($c$-[$ud$]) two-body bound state.
We extend the HAL QCD method to calculate the charm-diquark potential which reproduces the equal-time Nambu-Bethe-Salpeter wave function of the S-wave state ($\Lambda_c(\frac12^+)$).
The diquark mass is determined so as to reproduce the difference between the S-wave and the spin-orbit averaged P-wave energies, i.e.  the difference between the $\Lambda_c(\frac12^+)$ level and the average of the $\Lambda_c(\frac12^-)$ and the $\Lambda_c(\frac32^-)$ levels.
Numerical calculations are performed on a $32^3\times 64$ lattice with lattice spacing of $a \simeq 0.0907$ fm and the  pion mass of $m_{\pi}
\simeq 700$ MeV.
Our charm-diquark potential is given by the Coulomb+linear (Cornell) potential where the long range behavior is consistent with the charm-anticharm potential while the Coulomb attraction is considerably smaller.
This weakening of the attraction may be attributed to the diquark size effect.
The obtained diquark
mass is $m_D=1.273(44)$ GeV.
Our diquark mass lies slightly above the conventional estimates, namely the $\rho$  meson mass and twice the constituent quark mass $2m_N/3$.

\end{abstract}
\maketitle

\section{\label{Introduction}Introduction}
Understanding hadrons  as quark many-body  systems is one of  the most
important  themes  in  hadron  physics.
However, solving  many-body  problems  confronts  us  with  numerical
complexities even for 3-quark systems.
One way  to reduce the burden is  to introduce a  composite particle made
of    two   quarks called the  diquark
~\cite{GellMann,wilczek,jaffe}.
Then, for instance, a baryon  can be
considered as a bound state of a diquark and a quark.

Diquark  models~\cite{qD_model_santopinto,lichtenberg1,lichtenberg2}  that reduce the degrees of freedom in this way have  been  successful  in
accounting  for  a wide range of  problems,  e.g.   structures   and   reactions   of  hadrons   ~\cite{santopinto,   oettel, segovia,jido,Jido-size,chen}, 
exotic hadron levels~\cite{maiani,giannuzzi,giron} and the Regge trajectory in
the baryon sector~\cite{johnson}.
However,  the color confinement imposed by QCD forbids us to conduct direct experimental investigations on the diquark mass, quark-diquark interactions and so on.
They are merely inferred for practical calculations~\cite{mauro}.

Lattice  QCD  (LQCD)  Monte  Carlo  calculation, i.e. the  first-principle
calculation of QCD, makes it possible  to study various properties of diquarks.
For   instance, refs.\cite{diquark_mass,bi}  calculated the diquark mass.
However, it is still unclear whether the mass evaluated in these references for an isolated diquark is valid in a baryon.
Refs.\cite{diquark_size,francis}  calculated the spatial size of the diquark.
However, the interaction between a quark and a diquark is still unclear.

Recently, the HAL QCD  collaboration proposed a method to calculate hadron-hadron
potentials  from   the  equal-time  Nambu-Bethe-Salpeter   (NBS)  wave
function~\cite{10.3389/fphy.2020.00307,NN_Ishii,BB_Iritani} calculated by LQCD.
In 2011, Ikeda and Iida applied the HAL QCD method to
the quark-antiquark ($q\bar q$) system~\cite{Ikeda-Iida}.
Their $q\bar{q}$ potential was shown to behave as the well-known Cornell (Coulomb+linear) potential~\cite{Bali,Bali,Bali3,kinoshita}.
However, Ikeda and Iida  set the quark mass to half of the vector meson mass based on a naive constituent quark picture.
 Kawanai and Sasaki proposed a method to determine the quark mass in 2011~\cite{Kawanai-Sasaki,Kawanai-Sasaki2}.
 They did so by imposing a condition on the quark mass by making use of the spin-spin interaction between the quark and the anti-quark.
The quark mass calculated by the Kawanai-Sasaki method reproduced
the mesonic excitation spectrum with a satisfactory
accuracy~\cite{Nochi_eta_c,Kawanai-Sasaki2}.
These works together paved a way to evaluating the quark mass and the $q\bar q$ interaction potential which are applicable to mesons in general.

This paper extends the works on $q\bar q$ to the quark-diquark system, namely baryons.
To be specific, we consider the $\Lambda_{c}$ baryon as  a two-body system consisting of a
charm quark ($c$) and a [$ud$] scalar diquark ($D$).
However,  the Kawanai-Sasaki
method does not work in its original form for the  $\Lambda_{c}$ baryon.
This is because spin-less scalar diquark and the charm quark do not have spin-spin interaction.
Here, we  report  a  novel method to determine  the  diquark  mass, the quark mass and the quark-diquark interaction potential.
The condition we impose is that the so-determined quark-diquark potential and the diquark mass reproduce the energy difference between S-wave state ($\Lambda_c(\frac12^+)$) and the spin-orbit averaged P-wave states (average of $\Lambda_c(\frac12^-)$ and $\Lambda_c(\frac32^-)$).

This paper is organized as follows.
In  \Sect{Meson_formalism}, we  consider the  $q\bar q$
system to describe our formalism taking up the charm-anticharm ($c\bar c$) system as a specific example.
In \Sect{qD_formalism}, we apply our formalism to the $cD$ system
where the diquark mass is determined  by demanding it to reproduce the
P-wave excitation energy of the $\Lambda_c$ baryon.
In \Sect{setup}, we explain the  setup of our lattice QCD calculation,
i.e. the quark actions, gauge action and the LQCD parameters.
\Sect{Meson_Num} shows the  numerical results for the $c\bar  c$ system, namely the $c\bar c$ potential and the charm quark mass.
\Sect{cD_Num} shows the  numerical results  for the  $cD$ system, namely the $cD$ potential and the diquark mass.
 \Sect{ccbar_vs_cD} compares the $cD$ potential and the $c\bar c$ potential.
Finally, we summarize our  work and
give some future prospects in  Section \ref{Ketsuron}.

\section{\label{Meson_formalism}Formalism for $q\bar{q}$ sector}
In this section, we shall present our formalism for the $q\bar q$ system.
We focus on the charmonium which is a two-body system of a heavy charm quark ($c$) and an anti-charm quark ($\bar c$).
Such a system can be treated in the non-relativistic framework so that the interaction between the charm and anti-charm quarks can be expressed by a potential~\cite{kinoshita, Koma_Koma}.
Its application to other heavy quark systems, such as the bottomium, is straightforward.

\subsection{$c\bar c$ Numbu-Bethe-Salpeter wave function}
We start with the definition of the $c\bar{c}$  equal-time  Numbu-Bethe-Salpeter  (NBS)  wave   function $\phi_{\mathcal M}(\bm r)$ for the meson  state $\left| {\mathcal M}\right\rangle$ in the rest frame, namely
\begin{equation}
  \phi_{\mathcal M}(\bm r)
  \equiv
  \sum_{\alpha,\beta=1,2,3,4}
  \left\langle 0 \left|
  \bar c_{c,\alpha}(\bm x) \Gamma_{\mathcal M,\alpha\beta} c_{c,\beta}(\bm y)
  \right| {\mathcal M}
  \right\rangle,
  \label{eq:NBS-wave.implicit}
\end{equation}
where $|0\rangle$ is the QCD vacuum and $\bm r=\bm x - \bm y$ is the relative coordinate.
Here, the operator $c_{c,\alpha}(\bm x)$ denotes the charm quark field with color $\bm 3$ index
$c$ and spinor index $\alpha$.
The composite operator $\bar c_{c,\alpha}(\bm x) \Gamma_{\mathcal M,\alpha\beta} c_{c,\beta}(\bm y)$ annihilates the meson $\mathcal M$.
Contracting the charm and anti-charm quark operators with the Dirac matrix $\Gamma_{\mathcal M,\alpha\beta}$ specifies the parity and the spin of the annihilation operator.
The NBS wave function given by \Eq{eq:NBS-wave.implicit} is gauge dependent since the quark and the antiquark are located at different positions.
We fix the gauge to the Coulomb gauge in order to obtain the signal.
Representations of the meson $\mathcal M$ and the corresponding Dirac matrix $\Gamma_{\mathcal M,\alpha\beta}$ are summarized in \Table{table:gamma}.
For notational simplicity, we omit the spinor and color indices hereafter unless otherwise noted.
 
  \begin{table}
\caption{\label{table:gamma}  List   of  $J^{PC}$, spectroscopic
  notation $^{2S+1}L_J$,  and Dirac matrix $\Gamma_{\mathcal M}$ for
  the $c\bar{c}$ meson state $\mathcal M$.}
\begin{tabular}{c|ccccccccc}
  ${\mathcal M}$ & PS && V && S && AV && T \\
  \hline
 $J^{PC}$ & $0^{-+}$ && $1^{--}$ && $0^{++}$ && $1^{++}$ && $1^{+-}$ \\
 $^{2S+1}L_J$ & $^1S_0$ &\ & $^3S_1$-$^3D_1$ &\ & $^3P_0$ &\ & $^3P_1$ &\ & $^1P_1$ \\
  $\Gamma_{\mathcal M}$ & $\gamma_5$ && $\gamma_i$ && $I$ && $\gamma_5\gamma_i$ && $\gamma_i\gamma_j$
\end{tabular}
\end{table}

We define the $c\bar c $ potential $\hat V_{\mathcal M}$ by demanding the NBS wave function to satisfy the following Schr\"odinger equation in the region $E_\mathcal M < 2 M_{{\rm D_{meson}}}$, i.e. below the  pair creation threshold of a $D$ meson and a $\bar D$ meson~\cite{Ikeda-Iida, Kawanai-Sasaki}:
\begin{equation}
  \left(-\frac{\nabla^2}{2\mu} + \hat V_{\mathcal M}\right)
  \phi_{\mathcal M}(\bm r)
  =
  E_{\mathcal M}
  \phi_{\mathcal M}(\bm r),
  \label{eq:schroedinger}
\end{equation}
where $\mu=m_c/2$ is the reduced mass of the $c\bar c$ system and $E_{\mathcal M}=M_\mathcal M-2m_c$ is the binding energy where $m_c$ is the charm quark mass and $M_{\mathcal M}$ is the meson mass.
At this point, $m_c$ is undetermined.
This is because the quark cannot be isolated due to the color confinement prohibiting direct measurements of $m_c$.
Determination of $m_c$ will be discussed in the next subsection.
On the other hand, the meson mass $M_{\mathcal M}$ is extracted
from the temporal meson correlator where details will be addressed in \Sect{Meson_Num}.

It is shown in refs.\cite{Ikeda-Iida, Kawanai-Sasaki} that the potential operator $\hat  V_\mathcal M$  in \Eq{eq:schroedinger} is energy-independent and non-local, and acts on the wave function as an integral operator~\cite{NN_Ishii,10.3389/fphy.2020.00307}
\begin{equation}
\hat  V_\mathcal M \psi_\mathcal M(\bm r) =\int d^3r' V_\mathcal M(\bm r, \bm r')\psi_\mathcal M(\bm r').
\label{eq:non-local_pot}
\end{equation}
The derivative expansion~\cite{10.3389/fphy.2020.00307,Sugiura} $\hat V_\mathcal M(\bm r, \bm r')=V_\mathcal M(\bm r, \nabla) \delta^3(\bm r - \bm r')$ yields
\begin{widetext}
\begin{eqnarray}
  V_\mathcal M(\bm r,\nabla)
  =
  V_{0}(r)
  + V_{\sigma}(r)\bm s_1\cdot\bm s_2
  + V_{\rm TN}(r) S_{12}
  + V_{\rm LS}(r) \bm L\cdot \bm S
  + O(\nabla^2),
  \label{eq:derivative_expansion}
\end{eqnarray}
\end{widetext}
where  $V_{0}(r)$, $V_{\sigma}(r)$,  $V_{\rm TN}(r)$  and $V_{\rm  LS}(r)$
denote  the spin-independent,  spin-spin,  tensor and  spin-orbit
potentials,  respectively.
The operators $S_{12}  \equiv  4\left\{(\bm   r\cdot\bm  s_1)(\bm  r\cdot  \bm
s_2)/r^2 - \bm s_1\cdot\bm s_2/3\right\}$,
$\bm s_1$,  $\bm s_2$,
$\bm L \equiv  -i \bm r \times  \bm\nabla$ and
$\bm S\equiv  \bm s_1 + \bm  s_2$ denote the tensor  operator, the spin
operators  for  the charm  and  the  anti-charm quarks, the  orbital  angular
momentum operator and the total spin operator, respectively.
The higher order term $\mathcal O(\nabla^2)$ is neglected hereafter.
Recalling \Eq{eq:non-local_pot}, the integral operator $\hat V_\mathcal M$ in \Eq{eq:schroedinger} is replaced by $V_\mathcal M(\bm r,\nabla)$.

\subsection{Determination of $m_c$}
Since the charm quark mass in \Eq{eq:schroedinger} is undetermined,  the $q\bar q$ potential is also undetermined at this point. 
In place of the potential, we first define the {\it pre-potential} $\tilde{V}_\mathcal M(\bm r)$ by
\begin{equation}
  \tilde V_{\mathcal M}(\bm r)
  =
  \frac{\nabla^2 \phi_{\mathcal M}(\bm r)}{\phi_{\mathcal M}(\bm r)}.
  \label{eq:pre-potential}
\end{equation}
Recalling \Eq{eq:schroedinger},  we see that the
pre-potential is related to the potential as
\begin{equation}
\tilde{V}_\mathcal M(\bm r)\equiv m_c\left\{V_\mathcal M (\bm r, \nabla)-E_\mathcal M \right\}.
\label{eq:pre-potential_2}
\end{equation}
Hence, the pre-potential is the potential shifted by the binding energy $E_\mathcal M$ and factorized by $m_c$.
Explicit forms of the pre-potential for the PS, V, S, AV and T states (see \Table{table:gamma} for their representations) are obtained as follows.
First, $V_\mathcal M (\bm r, \nabla)$ in \Eq{eq:pre-potential_2} is replaced by its form given in \Eq{eq:derivative_expansion}.
Then, the operators $S_{12}$,
$\bm s_i$,
$\bm L $ and
$\bm S$ are replaced by their eigenvalues.
The results are, 
\begin{widetext}
\begin{eqnarray}
  \tilde V_{\rm PS}(r)
  &=&
  m_c\left(V_0(r) - (3/4) V_{\sigma}(r) - E_{\rm PS}\right)
  \nonumber\\
  \tilde V_{\rm V}(r)
  &\simeq&
  m_c\left(V_0(r) + (1/4) V_{\sigma}(r) - E_{\rm V}\right)
  \nonumber\\
  \tilde V_{\rm S}(r)
  &=&
  m_c\left(V_0(r) + (1/4) V_{\sigma}(r) - 4 V_{\rm TN}(r) - 2 V_{\rm LS}(r)
  - E_{\rm S}
  \right)
  \nonumber\\
  \tilde V_{\rm AV}(r)
  &=&
  m_c\left(V_0(r) + (1/4) V_{\sigma}(r) + 2 V_{\rm TN}(r) - V_{\rm LS}(r)
  - E_{\rm AV}
  \right)
  \nonumber\\
  \tilde V_{\rm T}(r)
  &=&
  m_c\left(V_0(r) - (3/4) V_{\sigma}(r) - E_{\rm T}\right).
  \label{eq:qqbar_pre-potentials}
\end{eqnarray}
\end{widetext}
Here, we have neglected the tensor potential in the V state originating from the small D-wave mixing~\cite{Nochi_eta_c}.

\subsubsection{The Kawanai-Sasaki method for $m_c$}
Before moving on, we  give  a  brief  review   of the  Kawanai-Sasaki  method which is one way to  determine
$m_c$. 
In this method, it is demanded that the  spin-spin interaction potential vanish at long distances,
which leads to
\begin{eqnarray}
  0
  &=&
  \lim_{r\to\infty}m_c V_{\sigma}(r)
  \\\nonumber
  &=&
  \lim_{r\to\infty}
  \left(
  \tilde V_{\rm V}(r)
  - \tilde V_{\rm PS}(r)
  - m_c \Delta E_{\rm hyp}
  \right),
\end{eqnarray}
where $\Delta E_{\rm hyp} \equiv E_{\rm V}  - E_{\rm PS} = M_{\rm V} -
M_{\rm PS}$. 
As a result, $m_c$ is uniquely determined as
\begin{eqnarray}
  m_{c}
  &\equiv&
  \frac1{\Delta E_{\rm hyp}}
  \lim_{r\to\infty}
  \left(
\tilde V_{\rm V}(r)
  - \tilde V_{\rm PS}(r)
  \right)\nonumber \\
  &=&
    \frac1{\Delta E_{\rm hyp}}
  \lim_{r\to\infty}
  \left(
  \frac{\nabla^2 \phi_{\rm V}(r)}{\phi_{\rm V}(r)}
  -
  \frac{\nabla^2 \phi_{\rm PS}(r)}{\phi_{\rm PS}(r)}
  \right)
\end{eqnarray}
where we have used \Eq{eq:pre-potential} to obtain the last line.
Note that the Kawanai-Sasaki method requires the spin-spin interaction potential $V_\sigma(r)$ to be finite at moderate distances.

\subsubsection{A new method to determine $m_c$}
We propose an alternative method  to determine $m_c$ without making use of the spin-spin
interaction potential.
Our strategy is to determine the value of $m_c$ so that the Schr\"odinger equation \Eq{eq:schroedinger} reproduces the P-wave excitation energy evaluated using meson masses.

We  focus  on the  spin-singlet sector,  i.e.  PS and  T states which are S-wave and P-wave of our interest.
The absence of the spin-orbit potential and the tensor potential in the T state makes the determination of $m_c$ simpler compared with the spin triplet sector (see \Appendix{appendix:m_q_spin_singlet} for details).
Indeed, it is shown from \Eq{eq:qqbar_pre-potentials} that the pre-potentials for the PS state and T state are related by
\begin{equation}
  \tilde V_{\rm  T} =  \tilde V_{\rm  PS} +\tilde E_{\rm  PS,T}.
  \label{eq:T-PS-pre-pot}
\end{equation}
Here, we have introduced the pre-energy $\tilde E_{\rm  PS,T}  \equiv  m_c(E_{\rm T}  -  E_{\rm PS})$.
Note that $\tilde E_{\rm  PS,T} =m_c\left(M_{\rm T} - M_{\rm PS}\right)$ follows from the definition of the binding energies $E_{\rm T}$ and $E_{\rm PS}$.
We see from \Eq{eq:T-PS-pre-pot} that the only difference between S-wave and P-wave equations is the centrifugal term in the kinetic energy.
Hence, we get the following radial part Schr\"odinger equations:
\begin{eqnarray}
  \left(-\frac{1}{r^2}\frac{d}{dr}\left(r^2\frac{d}{dr}\right)+ \tilde V_{\rm PS}(r)
  \right)
 \phi_{\rm PS}(r)
  &=&
  0
  \label{eq:pre-schroedinger}
  \nonumber\\
  \left(
  -\frac{1}{r^2}\frac{d}{dr}\left(r^2\frac{d}{dr}\right)+\frac{2}{r^2} + \tilde V_{\rm PS}(r)
  \right)
  \phi_{\rm T}(r)
  &=&
  \tilde E_{\rm PS,T}
  \phi_{\rm T}(r).
  \nonumber\\
  \label{eq:P-wave-excitation}
\end{eqnarray}
Now, we proceed as follows.
First, we calculate the NBS wave function for the PS state from LQCD using \Eq{eq:NBS-wave.implicit}.
 Next, we construct the pre-potential $\tilde  V_{\rm PS}(r)$ from the NBS wave function using \Eq{eq:pre-potential}. 
  Then, we calculate the excitation pre-energy $\tilde E_{\rm PS,T}$ by solving the lower of Eq's.(\ref{eq:P-wave-excitation}).
Finally, $m_c$ is given by
\begin{equation}
  m_{c}
  \equiv
  \frac{\tilde E_{\rm PS,T}}{M_{\rm T}-M_{\rm PS} }.
  \label{charm_mass}
\end{equation}

\section{\label{qD_formalism}formalism for quark-diquark sector}
Next, we formulate our method for the quark-diquark sector.
To be specific, we focus on the $\Lambda_c$ baryon considering it as a bound state formed by
a charm quark and a scalar $[ud]$ diquark ($D$) whose spin, parity,  isospin, and color representation are  $J^{P}=0^+$, $I=0$ and $\overline{\bm 3}$, respectively.
We  focus  on  the  S-wave  state ($\Lambda_c(\frac12^+)$)  and  the two P-wave  states ($\Lambda_c(\frac12^-)$  and
$\Lambda_c(\frac32^-)$) split by the spin-orbit interaction.
The spectroscopic notations for the $\Lambda_c$ states are summarized in \Table{table:spectroscopic-cD}.
\begin{table}[t]
  \caption{Spectrospcopic notations for $\Lambda_c$ states as the quark-diquark two-body bound states.}
  \begin{tabular}{c|ccc}
    $\Lambda_c(J^{P})$ & $\Lambda_c(\frac12^+)$ & $\Lambda_c(\frac12^-)$ & $\Lambda_c(\frac32^-)$
    \\
    \hline
    $^{2S+1}L_J$ & $^2S_{1/2}$ & $^2P_{1/2}$ & $^2P_{3/2}$
  \end{tabular}
  \label{table:spectroscopic-cD}
\end{table}

The outline of our method is as follows.
First, we define the $cD$ potential by demanding the $cD$ NBS wave function for the $\Lambda_c(\frac12^+)$ state to satisfy a Schr\"odinger equation. 
Next, the equation is solved to obtain the LS averaged P-wave excitation energy.
 Then, the diquark mass is determined by equating the P-wave excitation energy to the difference between $\Lambda_c(\frac12^+)$ energy and the spin-orbit average of $\Lambda_c(\frac12^-)$ and $\Lambda_c(\frac32^-)$ energies.
\subsection{cD NBS wave function}
Let us start with the $cD$ NBS wave function $\psi_{\Lambda_c(J^P); \alpha}(\bm r)$ for the $\Lambda_c(J^P)$ state given by
\begin{equation}
  \psi_{\Lambda_c(J^P); \alpha}(\bm r)
  \equiv
  \left\langle 0 \left|
  D_c(\bm x)
  c_{c,\alpha}(\bm y)
  \right| \Lambda_c(J^P)\right\rangle,
  \label{eq:charm-diquark_NBS}
\end{equation}
where $|\Lambda_c(J^{P})\rangle$ denotes the $\Lambda_c$ baryon state with its spin-parity denoted by
$J^{P}$.
The composite [$ud$] scalar-diquark operator $D_c(\bm  x)$ is defined by
\begin{equation}
  D_c(\bm x)
  \equiv
  \epsilon_{abc} u_a^T(\bm x) C\gamma_5 d_b(\bm x),
\end{equation}
where $u_a(\bm x)$ ($d_b(\bm x)$) is the field operator of the $u$ ($d$)  quark and $C\equiv i\gamma^2\gamma^0$ the charge conjugation matrix. 
The Levi-Civita symbol $\epsilon_{abc}$ is introduced to construct the color $\overline{\bm 3}$ diquark operator.
The $cD$ NBS wave function given by \Eq{eq:charm-diquark_NBS}  is gauge-dependent.
Thus, we fix the gauge before calculating the NBS wave function as in the case of the $c\bar c$ system.

We define the $cD$ potential $U(\bm r, \nabla)$ by demanding the equal-time NBS wave function to satisfy the following Schr\"odinger equation:
\begin{equation}
  \left(
  -\frac{\nabla^2}{2\mu_{cD}}
  +U(\bm r, \nabla)
  \right)
  \psi_{\Lambda_c(J^P)}(\bm r)
  =
  E_{\Lambda_c(J^{P})}
  \psi_{\Lambda_c(J^{P})}(\bm r),
  \label{eq:schrodinger_qD0}
\end{equation}
where $\mu_{cD}\equiv  m_cm_D/(m_c+m_D)$ is  the reduced  mass of
the $cD$ system and $E_{\Lambda_c(J^{P})}\equiv M_{\Lambda_c(J^{P})} -  (m_c+m_D)$ is the binding energy.
Here, $m_D$ is the diquark mass to be determined in the next subsection.
 The  mass  $M_{\Lambda_c(J^{P})}$ of 
$\Lambda_c(J^P)$ is usually extracted from the corresponding temporal lattice correlator, and will be discussed in \Sect{cD_Num}.

The quark-diquark  potential $U(\bm r, \nabla)$ is expressed in terms of the derivative expansion
\begin{equation}
  U(\bm r, \nabla)
  =
  U_{0}(r)
  +
  U_{\rm LS}(r)\bm L \cdot \bm s,
 \label{eq:derivative_expansion_qD}
\end{equation}
where  $U_{0}(r)$  and $U_{\rm  LS}(r)$  denote  the central  and  the
spin-orbit potentials, respectively.

\subsection{Determination of $m_{D}$}
As in the previous section, let us first define the pre-potential $\tilde U_{\Lambda_c(J^P)}$ by
\begin{equation}
  \widetilde U_{\Lambda_c(J^P)}(r)
  \equiv
  \frac{\nabla^2 \psi_{\Lambda_c(J^P)}(\bm r)}{\psi_{\Lambda_c(J^P)}(\bm r)}.
  \label{cD_pre-pot}
\end{equation}
Substituting \Eq{eq:derivative_expansion_qD} into \Eq{eq:schrodinger_qD0} and rearranging with \Eq{cD_pre-pot}, we arrive at the following equations for $\Lambda_c(\frac12^+)$, $\Lambda_c(\frac12^-)$ and $\Lambda_c(\frac32^-)$ states:
\begin{widetext}
\begin{eqnarray}
  \left(
  -\nabla^2 + \widetilde U_{0}(r)
  \right)
  \psi_{\Lambda_c(\frac12^+)}(\bm r)
  &=&
  0
      \label{eq:p_split}
  \nonumber\\
  \left(
  -\nabla^2 + \widetilde U_{0}(r) - \widetilde U_{\rm LS}(r)
  \right)
  \psi_{\Lambda_c(\frac12^-)}(\bm r)
  &=&
  2 \mu_{cD}
  \left(
  E_{\Lambda_c(\frac12^-)} - E_{\Lambda_c(\frac12^+)}
  \right)
  \psi_{\Lambda_c(\frac12^-)}(\bm r)
  \nonumber\\
  \left(
  -\nabla^2 + \widetilde U_{0}(r) + \frac1{2} \widetilde U_{\rm LS}(r)
  \right)
  \psi_{\Lambda_c(\frac32^-)}(\bm r)
  &=&
  2 \mu_{cD}
  \left(
  E_{\Lambda_c(\frac32^-)} - E_{\Lambda_c(\frac12^+)}
  \right)
  \psi_{\Lambda_c(\frac32^-)}(\bm r)
  \label{eq:pre-shrodingers}
\end{eqnarray}
\end{widetext}
where $\widetilde U_{0}(r) \equiv 2\mu_{cD} (U_{0}(r)-E_{\Lambda_c(\frac12^+)})$ and $\widetilde U_{\rm LS}(r) \equiv 2\mu_{cD} U_{\rm LS}(r)$.
Here, the eigenvalues of the $\bm L\cdot \bm s$ operator are used explicitly, namely $0$,  $-1$  and  $1/2$
 for $\Lambda_c(\frac12^+)$, $\Lambda_c(\frac12^-)$
and  $\Lambda_c(\frac32^-)$, respectively.
  
The P-wave pre-energy splits into the following two due to $\tilde U_{\rm LS}(r)$:
\begin{widetext}
\begin{eqnarray}
  \widetilde E_{\rm PW}
  - \left\langle \widetilde{U}_{\rm LS} \right\rangle_{\rm PW}
  &\simeq&
  2\mu_{cD}\left(E_{\Lambda_c(\frac12^-)} - E_{\Lambda_c(\frac12^+)}\right)
  =
  2\mu_{cD}\left(M_{\Lambda_c(\frac12^-)} - M_{\Lambda_c(\frac12^+)}\right)
  \nonumber \\
  \widetilde E_{\rm PW}
  + \frac1{2} \left\langle \widetilde{U}_{\rm LS}\right\rangle_{\rm PW}
  &\simeq&
  2\mu_{cD}\left(E_{\Lambda_c(\frac32^-)} - E_{\Lambda_c(\frac12^+)}\right)
  =
  2\mu_{cD}\left(M_{\Lambda_c(\frac32^-)} - M_{\Lambda_c(\frac12^+)}\right),
  \label{eq:LS_multiplet}
\end{eqnarray}
\end{widetext}
where $\widetilde E_{\rm PW}$ is the difference between the pre-energy of the S-wave and the spin-orbit averaged P-wave, i.e. the P-wave eigenvalue of $  -\nabla^2 + \widetilde U_{0}(r)$ in the first equation of Eq's(\ref{eq:pre-shrodingers}).
The symbol $ \left\langle \widetilde{U}_{\rm LS} \right\rangle_{\rm PW}$  denotes the expectation value of $\widetilde{U}_{\rm LS}(r)$ with respect to the P-wave.
We have used the definition of the binding energy $E_{\Lambda_c(J^{P})}$ to get the RHS of each equation in Eq's(\ref{eq:LS_multiplet}).
Solving Eq's(\ref{eq:LS_multiplet}) with respect to $\mu_{cD}$, we arrive at
\begin{equation}
  \mu_{cD}
  =
  \frac{3}{2}\frac{\widetilde E_{\rm PW}}{M_{\Lambda_c(\frac12^-)} + 2M_{\Lambda_c(\frac32^-)} - 3M_{\Lambda_c(\frac12^+)}}.
  \label{eq:mu_cd}
\end{equation}
Once $\mu_{cD}$ is determined, the diquark mass $M_D$ is subsequently determined by
\begin{equation}
  M_{D}
  =
  \frac{\mu_{cD} m_c}{m_c - \mu_{cD}}.
  \label{eq:m_d}
\end{equation}

\section{\label{setup}Lattice QCD setup}
We use a set of QCD configurations generated by the  PACS-CS collaboration~\cite{pacs_config} using the Iwasaki gauge action~\cite{iwasaki1}  and the $\mathcal{O}(a)$-improved Wilson quark action~\cite{Improved_Wilson}.
The set has 399 of $2+1$ flavor QCD configurations generated on a $V\times T = 32^3 \times 64$ lattice at $\beta=1.90$ with the clover coefficient $c_{{\rm SW}}=1.715$ and the hopping parameters given by $\kappa_{u,d}=0.13700$  for the $u,d$ quarks and $\kappa_s=0.13640$ for the $s$ quark~\cite{pacs_config} .
The lattice spacing is $a \simeq 0.0907$ fm.
The statistical errors are estimated by the standard Jackknife method throughout this paper.

 The Iwasaki action for the gauge field is given by
 \begin{equation}
 S_{{\rm Iwasaki}}
 \equiv
 \frac{1}{{\it g}^2}
 \left\{
 c_0\sum_{{\rm plaquette}} {\rm Tr}[U_{{\rm plaq}}]
 +
 c_1 \sum_{{\rm rectangle}} {\rm Tr}[U_{{\rm rect}}]
 \right\}.
 \label{Iwasaki_action}
 \end{equation}
 Here, $U_{{\rm rect}}$ denotes the $1\times 2$ rectangle loop of the link variable.
 The constants are set to $c_0=3.648$ and $c_1=-0.331$~\cite{iwasaki1}.
 Such choice of the coefficients mitigates the discretization errors to $\sim\mathcal O(a^4)$.
 
The $\mathcal{O}(a)$-improved Wilson quark action $S_{{\rm quark}}$ for the light dynamical quarks ($u,d,s$) is given by
\begin{eqnarray}
S_{{\rm quark}}
\equiv
&&\sum_{q=u,d,s}\sum_{x}\left[
\bar q(x) q(x)
\right.\nonumber\\
&&
-
\kappa_q c_{{\rm SW}} \sum_{\mu,\nu}\frac{i}{2}\bar q(x) \sigma_{\mu\nu}F_{\mu\nu}(x)q(x) \nonumber \\
&-&
\kappa_q\sum_{\mu}\left\{
\bar q(x) (1-\gamma_\mu)U_{\mu}(x)q(x+\mu)\right.\nonumber \\
&+&
\left.\left.
\bar q(x) (1+\gamma_\mu)U^\dagger_{\mu}(x-\mu)q(x-\mu)\right\}\right].
\label{eq:improved_quark}
\end{eqnarray}
 Certain choice of the coefficient $c_{{\rm SW}} $ reduces the discretization errors to $\sim \mathcal O(a^2)$.
The action \Eq{eq:improved_quark} is used to calculate the $u$ quark and $d$ quark propagators.

 If we apply the $\mathcal{O}(a)$-improved Wilson quark action designed for light quarks to the heavy charm quark, it yields systematic errors of order $\sim \mathcal O(am_c)$ ~\cite{RHQ_on_LQCD}.
Therefore, we use the relativistic heavy quark action (RHQ) ~\cite{RHQ_on_LQCD,Namekawa_charm} with systematic errors reduced to the same order as the $\mathcal{O}(a)$-improved Wilson quark action.
 The RHQ action $S_{{\rm RHQ}}$ is given by
\begin{eqnarray}
S_{{\rm RHQ}}
\equiv
&&\sum_{x}\bar Q(x) Q(x) \nonumber \\
&&-\kappa_Q \sum_{x,y}\bar Q(x) \nonumber\\
&&\Biggl[\left.\sum_i 
\left(
(r_s-\nu \gamma_i)
U_i(x)\delta_{x+i,y}
+
(r_s+\nu \gamma_i)U^\dagger_i(x)\delta_{x,y+i}\right)\right. \nonumber \\
&&+(r_t-\nu\gamma_4)U_{\hat 4}(x)\delta_{x+\hat 4,y}
+(r_t+\nu\gamma_4)U^\dagger_{\hat 4}(x)\delta_{x,y+\hat 4}\nonumber \\
&&+\left.c_B \sum_{i,j}F_{ij}(x)\sigma_{ij}+c_E \sum_i F_{i4}(x)\sigma_{i4}\right]Q(y),\nonumber \\
\end{eqnarray}
where $\hat 4$ denotes the unit vector in the temporal direction  and $Q(x)=c(x)$ denotes the charm quark field.
 The parameter $r_t$ is set to $1$ while other parameters $r_s,c_B,c_E$ and $\nu$ are taken from Ref.~\cite{Namekawa_charm} in \Table{RHQ_params}.

\begin{table}
		\caption{\label{RHQ_params}Summary of the parameters used for the RHQ action for the charm quark. 
		The values are taken from Ref.~\cite{Namekawa_charm}.}
	\begin{center}
		\begin{tabular}{c c c c c }
		\hline
  	  	$\kappa_c$  & $\nu$         & $r_s$           & $c_B$          & $c_E$\\
		\hline 
		0.10959947 & 1.1450511 & 1.1881607 &  1.9849139 & 1.7819512\\
		\hline
   	 	\end{tabular}
		\label{table:Lattice_setup_charm}
	\end{center}
\end{table}

Since we use gauge invariant source and sink operators, the gauge is fixed to the Coulomb gauge.
This is done by minimizing the functional $F[U]\equiv\sum_{x}\sum_{i=1}^3\left\{{\rm Tr}\left[{\rm Re} U_i(x)\right]\right\}$~\cite{gattringer}.
This gauge fixing yields discretization errors $\sim O(a^2)$.

\section{\label{Meson_Num}Numerical results for $c\bar c$ system}
\subsection{Meson mass}
Let us start with the meson two-point  correlator defined by
\begin{equation}
  C^{\mathcal M}(t)
  \equiv
  \frac1{V}
  \sum_{\bm x}
  \left\langle 0 \left|
  T\left[ O^{\mathcal M}(\bm x,t_{{\rm sink}}) O^{\mathcal M\ \dagger}(t_{{\rm src}}) \right]
  \right| 0 \right\rangle,
  \label{eq:meson_corr}
\end{equation}
where $V$ is the lattice volume and $t\equiv t_{{\rm sink}}-t_{{\rm src}}$.
 Here, $ O^{\mathcal  M}=\bar c \Gamma_\mathcal M c$ such that $O^{\mathcal  M}(\bm x, t_{{\rm sink}})$ denotes the meson point sink operator and $O^{\mathcal  M \ \dagger}(t_{{\rm src}}) $ denotes the wall source.
The Dirac matrix $\Gamma_\mathcal M$ is chosen accordingly from \Table{table:gamma}.

The time-reversal symmetry $C^{\mathcal M}(t)=C^{\mathcal M}(T-t)$ is used to reduce the statistical  noise for the meson correlator. 
Also, the statistics improves as the data calculated for 16 different source points $t_{{\rm src}}=0,4,8,\cdots,60$ are averaged.
Moreover, the signals of the vector and the axial-vector meson correlators are improved by averaging with respect to the 3-dimensional lattice rotation.


In this work, we employ the periodic  boundary condition with respect to $t$.
Thus the correlator \Eq{eq:meson_corr} acquires the cosh form.
Thus, the effective mass $M_{\rm eff}(t)$  is the solution of the following equation
\begin{equation}
  \frac{C_{\mathcal M}(t + 1)}{C_{\mathcal M}(t)}
  =
  \frac{\cosh\left(M_{\rm eff}(t)\cdot(t + 1 - T/2)\right)}
  {\cosh\left(M_{\rm eff}(t)\cdot(t - T/2)\right)},
\end{equation}
where $T$ denotes the temporal extension of the lattice.

\begin{figure}[h]
 \begin{minipage}[b]{\hsize}
  \centering
\includegraphics[width=\textwidth]{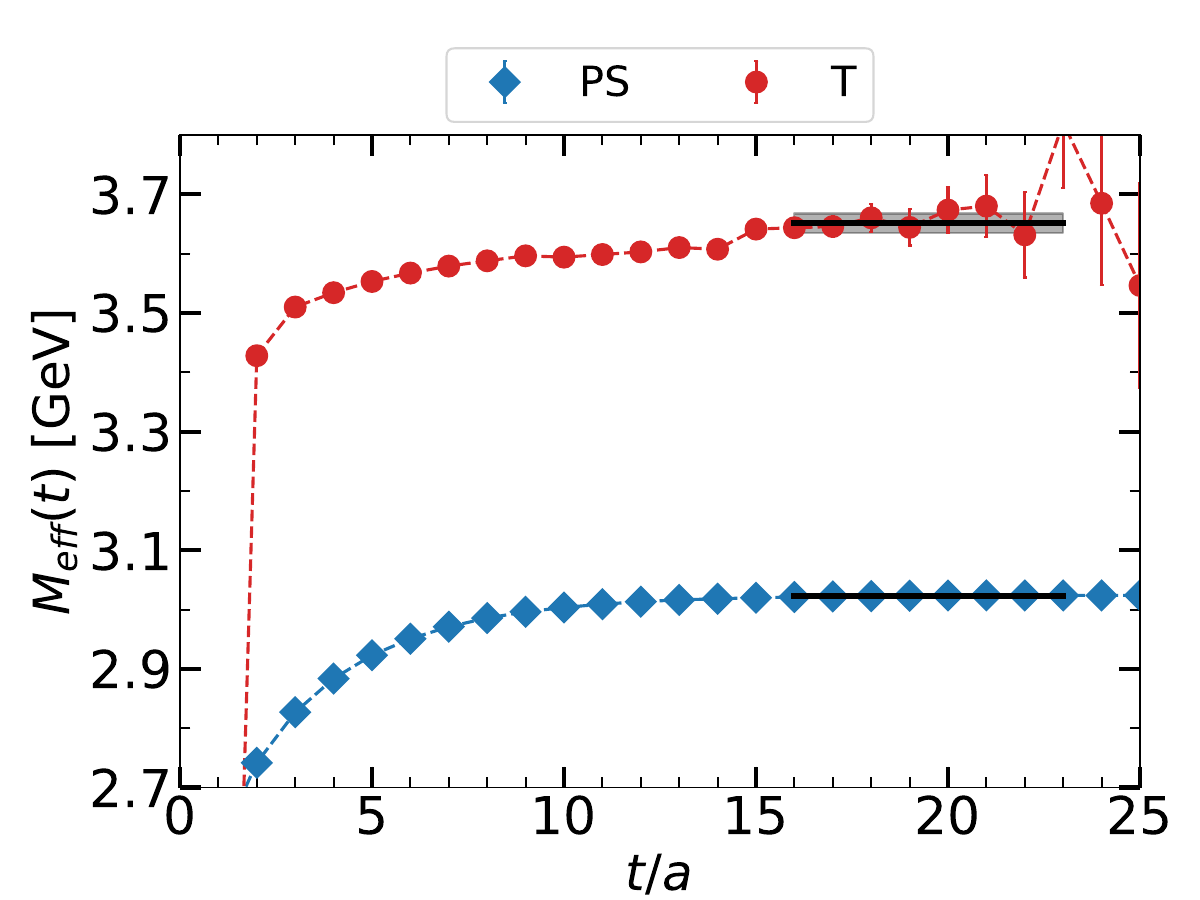}%
 \end{minipage}
 \begin{minipage}[b]{\hsize}
  \centering
\includegraphics[width=\textwidth]{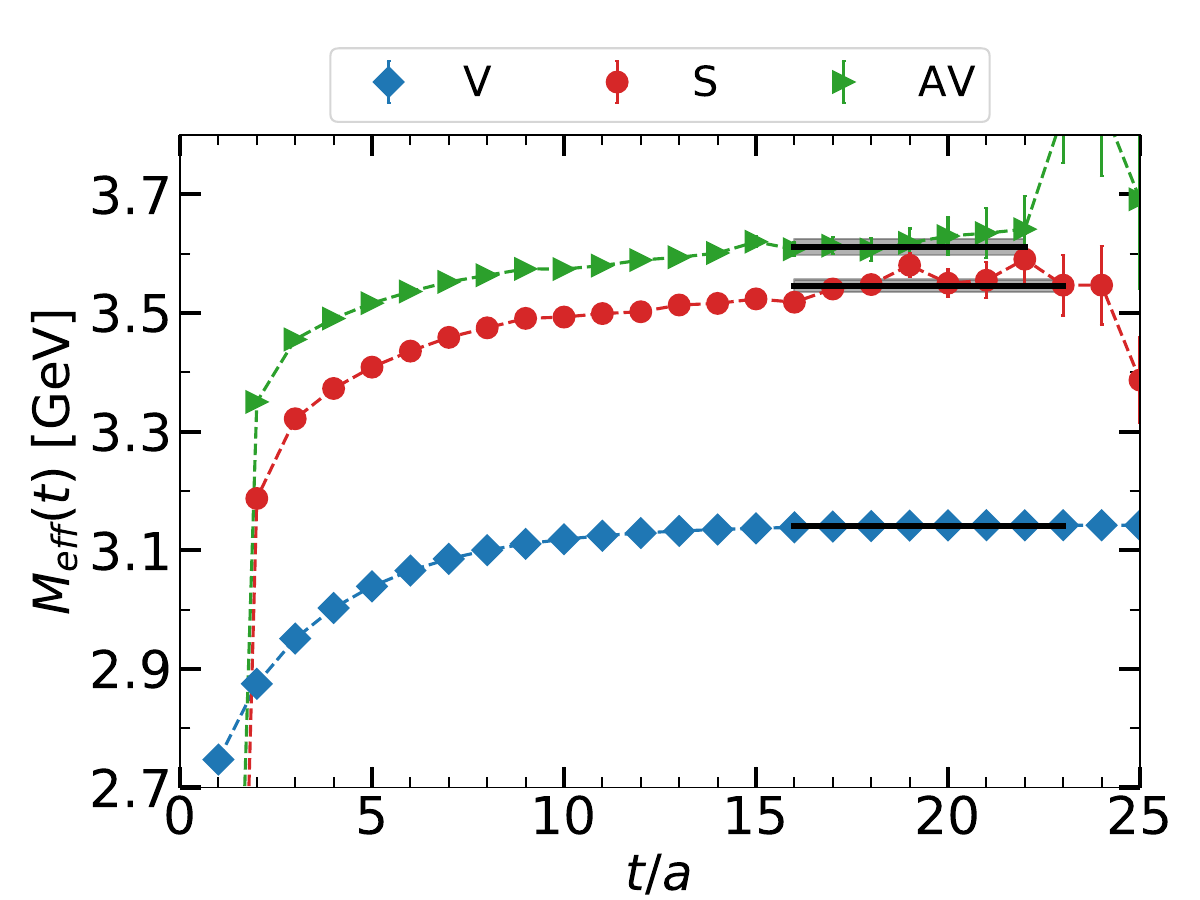}%
 \end{minipage}
\caption{Effective mass plots for the spin singlet (top) and triplet (bottom) sectors.
  The solid lines denote the average of the meson mass $M_\mathcal M$ over the fit range.
  The shaded areas denote the statistical errors estimated by the Jackknife method.\label{ccbar_EFM}}
\end{figure}

\Fig{ccbar_EFM} plots the meson effective mass $M_{\rm eff}(t)$ for PS, V , S, AV, and T states.
Then, the meson mass is extracted by fitting the correlator to $f_{\rm  fit}(t) \equiv  A\cosh(M(t - T/2))$ in each plateau region.
The fit is adequately carried out for each channel as shown in the figure with chi-square $\chi^2/N_{{\rm d.o.f}} < 1.0$, where $N_{{\rm d.o.f}}$ is the number of degrees of freedom in the fit.
The meson masses are summarized in \Table{tab:meson_mass_table}.
From the obtained meson masses, we get the hyperfine splitting $\varDelta E_{{\rm hyp}} = M_{\rm {V}} - M_{{\rm PS}} = 0.118 (1)$ GeV and the P-wave excitation energy for the spin singlet sector 
$E_{{\rm PS,T}} = M_{{\rm T}} - M_{{\rm PS}} = 0.628 (16)$ GeV.

For \Eq{eq:schroedinger} to be valid, the energy of the $c\bar c$ state must be below $2M_{{\rm D-meson}}$.
To check this condition, we apply the same procedure to the $D$ meson.
The result is $M_{{\rm D-meson}}  =  1.998(1)$ GeV, thus each $c\bar c$ states in \Table{tab:meson_mass_table} are bound.
For use in \Sect{cD_Num}, we also calculate the $\rho$ meson and the pion mass.
The results are $M_{\rho}  =  1.098(5)$ GeV and $M_\pi = 0.700(1)$ GeV.
\begin{table}[h]
\caption{\label{tab:meson_mass_table}$c \bar c$ meson masses. The fit ranges are shown in the last column.}
		\begin{tabular}{ c c c } 
		\hline\hline
		states          &  Mass [GeV] &  fit range \\ \hline
		PS       &  3.023 (1) & $16 \leq t/a \leq 23$ \\
		V        &  3.141 (1) & $16 \leq t/a \leq 23$\\
		S &  3.546 (11)& $16 \leq t/a \leq 23$ \\
		AV &  3.611 (14) & $16 \leq t/a \leq 22$ \\
		T       &  3.651 (16) & $16 \leq t/a \leq 23$\\
		\hline\hline
		\end{tabular}
\end{table}

\subsection{$c\bar{c}$ NBS wave function \label{Numerical_results}}
Next,  we consider the $c\bar c$ four-point correlator ~\cite{Ikeda-Iida,Kawanai-Sasaki,Kawanai-Sasaki2,Nochi_eta_c} in the rest frame
\begin{eqnarray}
    C_{\mathcal M}(\bm r, t)
  \equiv
  \frac1{V}
  \sum_{\bm \Delta}
  &
   \nonumber \\
   &
  \left\langle 0 \left|
  T\left[
    \bar{c}_c(\bm r + \bm \Delta, t_{{\rm sink}})
    \Gamma_{\mathcal M}
    c_c(\bm \Delta, t_{{\rm sink}})
    \right.
     \right.
      \right.
    \nonumber \\
    &
    \left.
    \left.
    \left.
    \times
    \mathcal{O}^{\mathcal M\ \dagger}(t_{{\rm src}})
    \right]
  \right| 0 \right\rangle.
  \label{ccbar-fourpoint}
\end{eqnarray}
Specifically, we calculate the correlator for the PS  and V states, i.e. for $\Gamma_\mathcal M = \gamma_5$ and $\Gamma_\mathcal M = \gamma_i$.
Here, the signal of the V state NBS wave function is improved by averaging the vector components symmetric with respect to the 3-dimensional lattice rotation.

Without loss of generality, we restrict ourselves to the $t>0$ region.
The four-point function $C_{\mathcal M}(\bm r, t)$ is spectrally decomposed as 
\begin{equation}
  C_{\mathcal M}(\bm r, t)
  =
  \sum_{n} a_n\phi_{\mathcal M}^{(n)}(\bm r) e^{-M_n t}
\end{equation}
where $\phi_{\mathcal M}^{(n)}(\bm r)$,  $E_n$ and $a_n \equiv \langle
n|\mathcal{O}^{\mathcal M\ \dagger}(0)|0\rangle$ denote  the equal-time NBS wave
function,  the energy  and the  overlap  for the  $n$-th
excited state $\left|n\right\rangle$, respectively.
The energy and the NBS wave function for $n=0$ correspond to the mass  and the wave function of the meson state ($\eta_c$ for PS and $J/\psi$ for V), respectively.
In the large $t$ region, the ground state $\left|0\right\rangle$ becomes dominant as $C_{\mathcal M}(\bm r, t)\sim a_0 e^{-M_0 t}\phi_{\mathcal M}^{(0)}(\bm r)$.
From now on, we focus on such a region.
Next, we project the NBS wave function to the $A_1$ representation~\cite{lusher} as
\begin{equation}
  \phi_{\mathcal{M}}(\bm r)
  =
  \frac1{48}
  \sum_{g \in O_h}
  \phi_{\mathcal{M}}(g^{-1} \bm r),
  \label{eq:A1+}
\end{equation}
where $O_h$ denotes the cubic group which has 48 elements (see \Appendix{App:cubic}).
This $ \phi_{\mathcal{M}}(\bm r)$ approximately represents the S-wave.
However, there are small contributions from the states with angular momentum $L\geq4$ that are not removed by the  $A_1$ projection~\cite{Ikeda-Iida,Ishizuka,Ishizuka2}.

\begin{figure}[h]
 \begin{minipage}[b]{\hsize}
  \centering
\includegraphics[width=\textwidth]{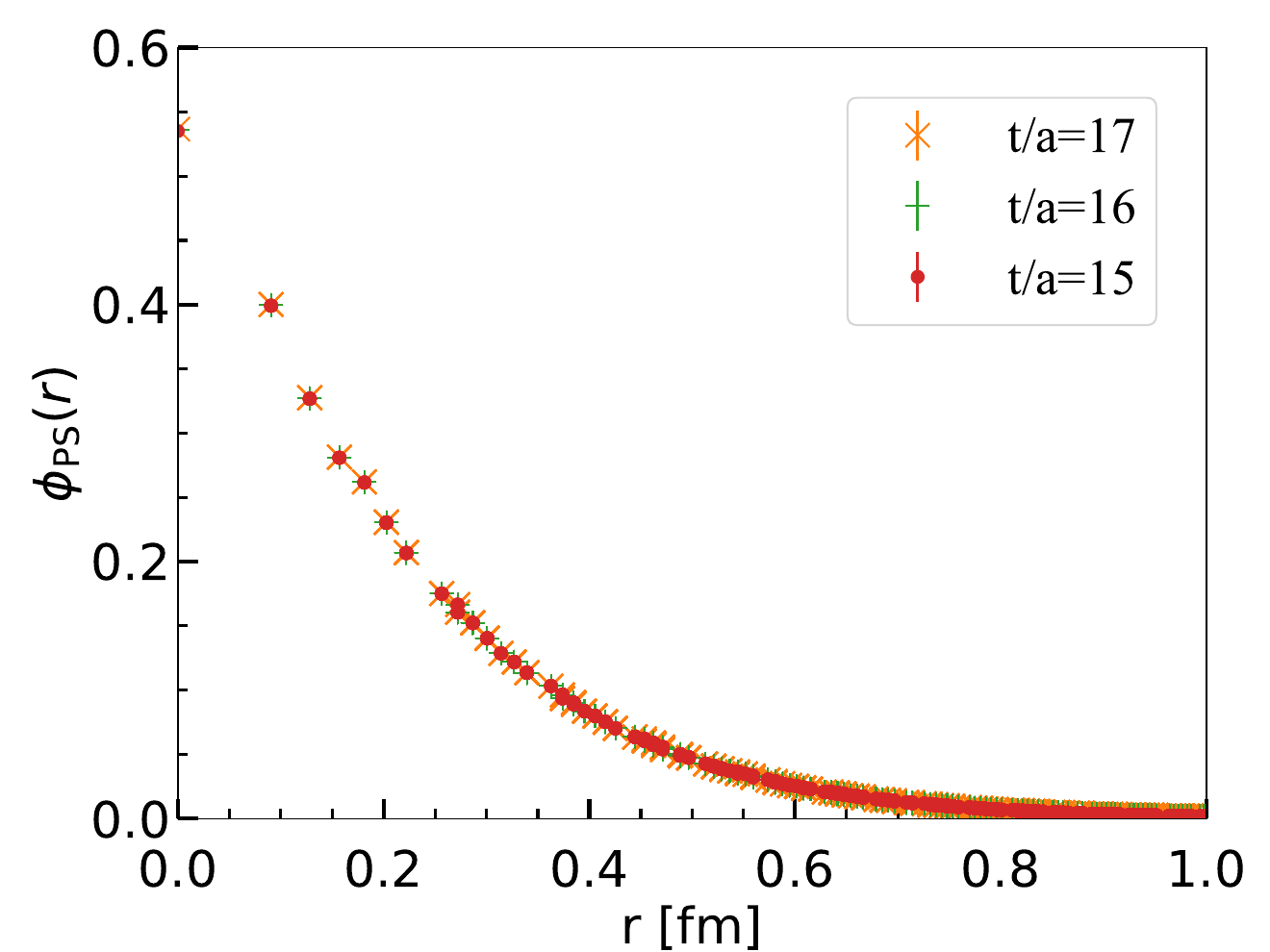}%
 \end{minipage}
 \begin{minipage}[b]{\hsize}
  \centering
\includegraphics[width=\textwidth]{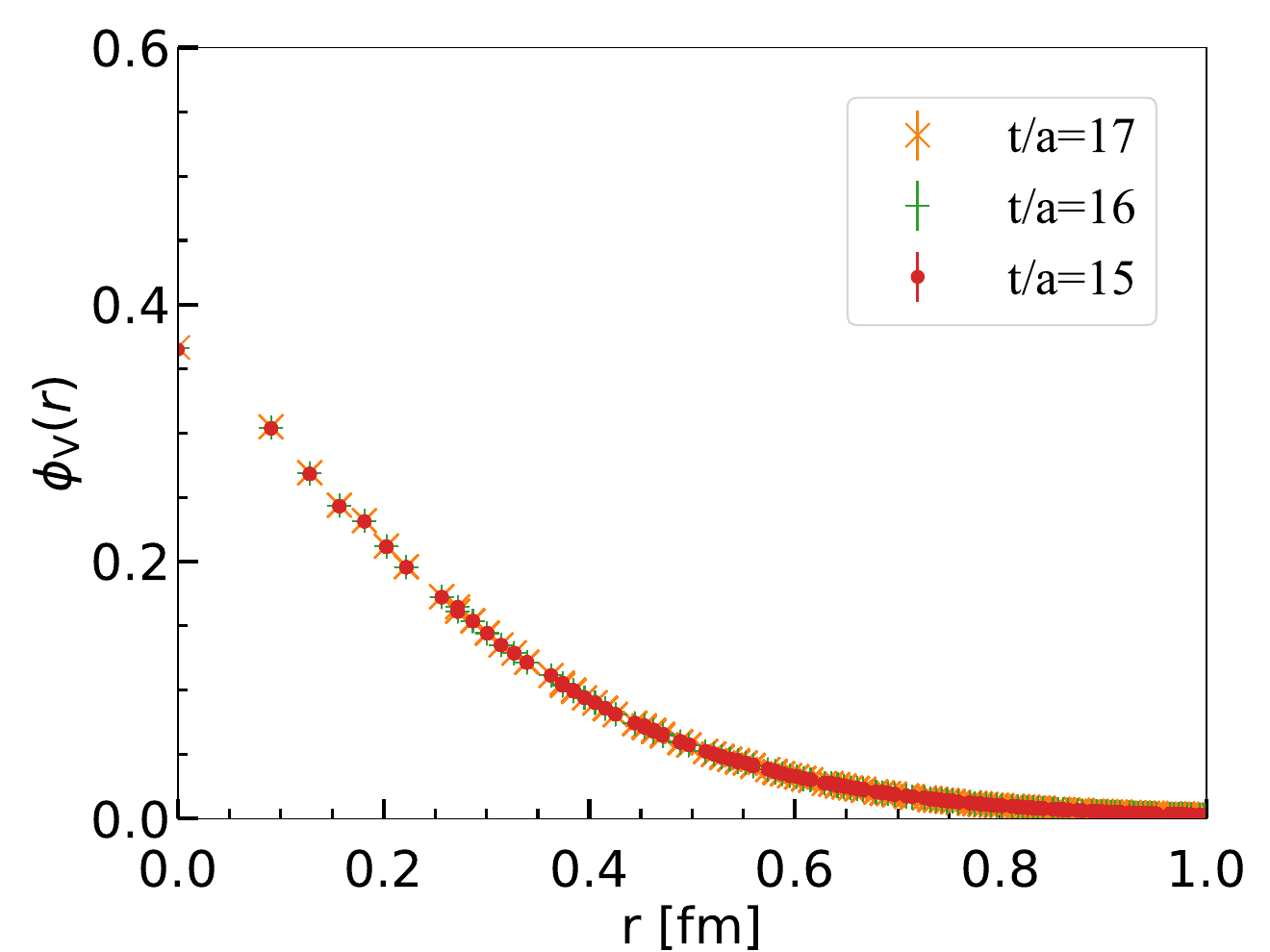}%
 \end{minipage}
\caption{NBS function for the PS (top) and V (bottom) states in $15 \geq t/a \geq 17$.
The NBS wave function is normalized as $\sum_{r \in V} r^2\phi_{\mathcal{M}}^2(\bm r) = 1$.
\label{ccbar_WF}}
\end{figure}

\Fig{ccbar_WF} shows the  NBS wave function for the PS state and that for the V state in several representative time slices.
We see that the NBS wave functions for $t/a=15,16,17$ fall on top of each other.
This indicates that the wave function in each time slice has attained its limit $\phi_\mathcal M (\bm r)$.
We focus on the wave function at $t/a=16$ since the wave functions all coincide in the time region $t/a \geq 15$.
Moreover, we focus on the region $0\leq r \leq 1.0$ fm since the NBS wave function is essentially zero beyond 1.0 fm.
Note that the NBS wave function has slight rotational symmetry breaking in the short range region.
Indeed, the wave function is slightly ``jagged'' near origin, as shown in \Fig{ccbar_WF}, which is an indication of  the breaking.

This symmetry breaking is due to the discretization errors. 
Possible sources of errors are the quark actions, the gauge action and the gauge fixing. 
Since the discretization errors of the gauge action are reduced by an optimal choice of the coefficients, the quark action and the Coulomb gauge fixing will remain to be the main sources of the rotational symmetry breaking. 

\subsection{Pre-potential for $c\bar{c}$}
Next, the pre-potential is evaluated from the NBS wave function according to \Eq{eq:pre-potential}.
Here, the Laplacian is numerically evaluated by the standard nearest-neighbor differentiation
\begin{equation}
\nabla^2 \phi_{\mathcal {M}}(\bm r)
=
\sum_{\hat \mu=\hat i_x, \hat i_y, \hat i_z}\left\{\phi_{\mathcal {M}}(\bm r+\hat \mu)+\phi_{\mathcal {M}}(\bm r-\hat \mu)\right\}-6\phi_{\mathcal {M}}(\bm r),
\end{equation}
where $\hat \mu$ denotes the unit vector in the positive $\mu$-direction and $\hat i_x$, $\hat i_y$ and $\hat i_z$ denote the unit vectors in $x$, $y$ and $z$ directions, respectively.

\begin{figure}[h]
\begin{center}
\includegraphics[width=0.48\textwidth]{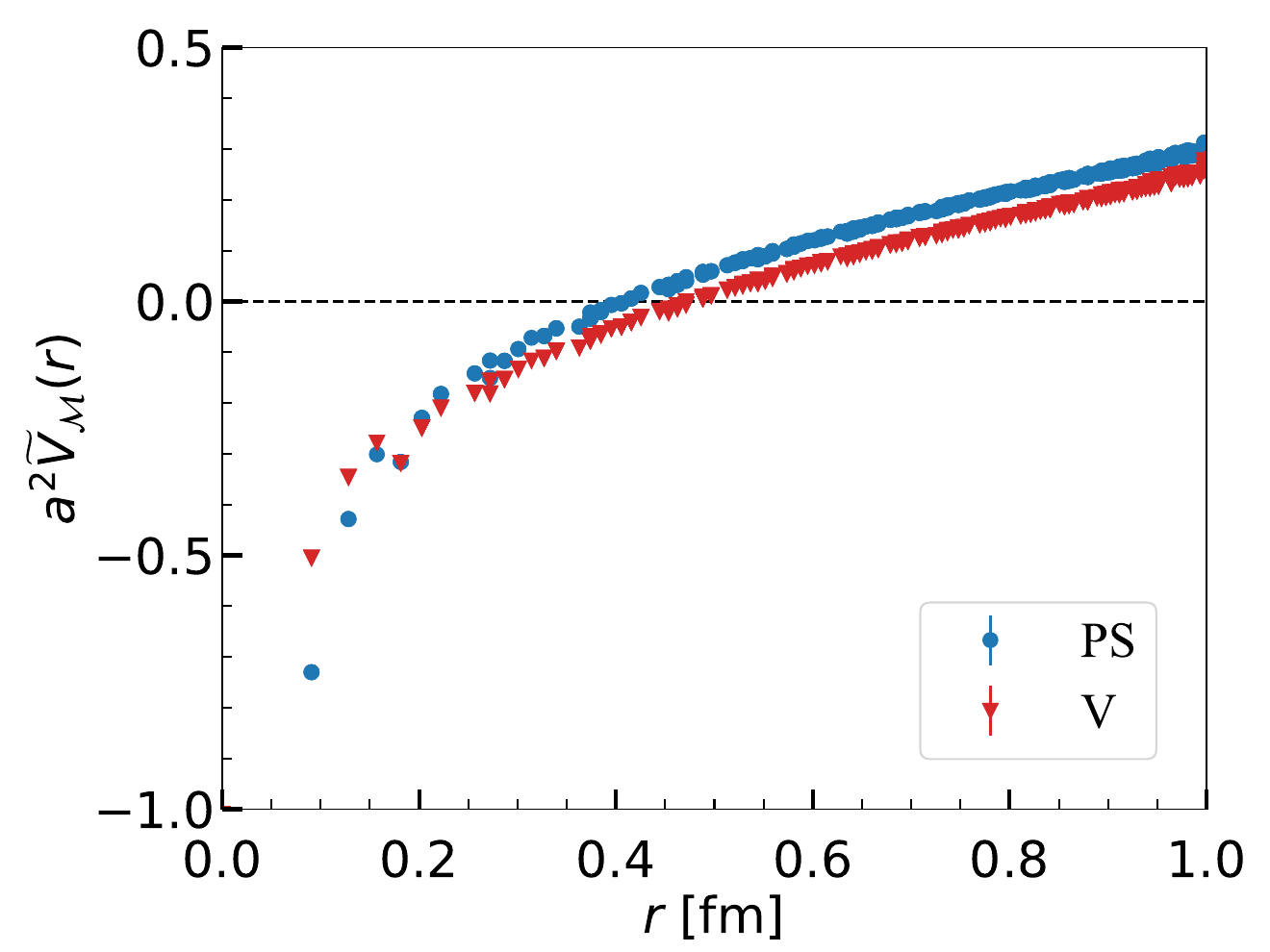}%
\caption{The pre-potentials for the PS (blue dots) and V (red triangles) states.
\label{ccbar_pre-pot}}
\end{center}
\end{figure}

\Fig{ccbar_pre-pot} shows the PS pre-potential and the V pre-potential.
These pre-potentials behave roughly as the well-known Coulomb+linear (Cornell) potential as expected.
However, we see in \Fig{ccbar_pre-pot} that  the pre-potential is jagged in the short range region due to the broken rotational symmetry passed on from the NBS wave function.

Next, we fit the pre-potential.
Using \Eq{eq:qqbar_pre-potentials}, the pre-potential is separated into the spin-independent part $\tilde{V}_0(r)$ and the spin-spin part $\tilde{V}_{\sigma}(r)$ as
\begin{eqnarray}
\tilde V_0(r)&=&m_q\left\{V_0(r)-\bar{E}_{1S}\right\}= \frac{3}{4}\tilde V_{\rm V} (r) +\frac{1}{4}\tilde V_{\rm PS}(r)
\nonumber\\
\tilde V_\sigma(r)&=&m_q\left\{V_\sigma(r)-\varDelta E_{hyp}\right\}=\tilde V_{\rm V} (r) -\tilde V_{\rm PS}(r),
\nonumber\\
\label{eq:spin-separate}
\end{eqnarray}
where $\bar{E}_{1S}=\frac{3}{4}E_{\rm V}+\frac{1}{4}E_{\rm PS}$.
We fit the spin-independent part to the Cornell+log function $ V^{{\rm fit}}_0(r) $ given by
\begin{equation}
 V^{{\rm fit}}_0(r) 
 =
- A/r
+
B r
+
C\log(r/a)
+
v_0.
\end{equation}
Here, the Cornell term is expected both phenomenologically~\cite{kinoshita} and theoretically~\cite{Bali} for heavy quark systems.
The logarithmic term $C\log(r/a)$ is a correction from the finite  quark mass~\cite{Koma_Koma,Koma_Koma2,log-pot}  where the lattice spacing $a$ is introduced to set the argument dimensionless.
The spin-spin part is fit to the 2-Gaussian function $V^{\rm fit}_{\sigma}(r)$ given by
\begin{equation}
  V^{\rm fit}_{\sigma}(r)
  =
  A_1\exp(-B_1r^2) + A_2\exp(-B_2r^2) + {v_\sigma}.
  \label{eq:2Gauss}
\end{equation}

\begin{figure*}[ht]
\includegraphics[width=\textwidth]{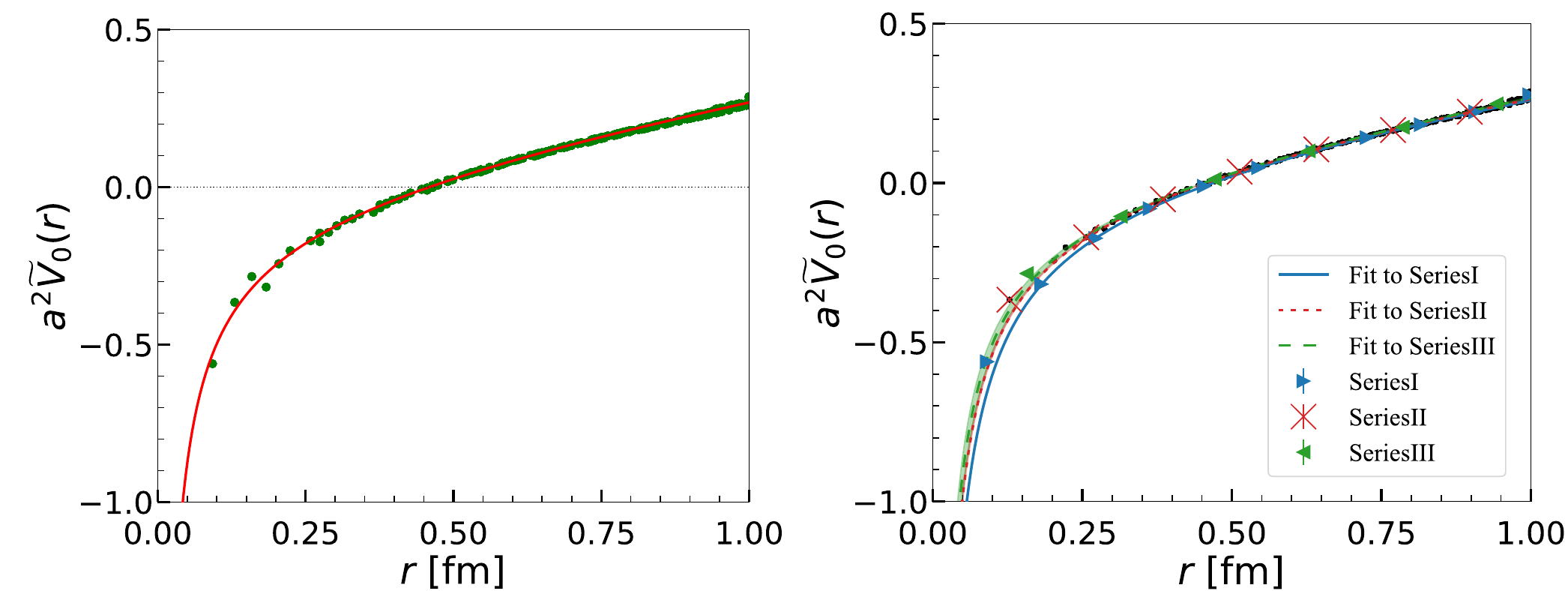}%
\caption{\label{ccbar_fit_ALL}Spin-independent part of the pre-potential (dots) fit to the Cornell+log function (solid line).
(left) Fit to ALL and (right) fit to Series I, \rmII and \rmIII.
The shaded area around each line denotes statistical errors.}
\end{figure*}

First, we fit the spin-independent part of the pre-potential to the Cornell+log function.
We fit in the range $1 \leq r/a \leq11$ ($0.09\lessapprox r \lessapprox 1.0$ fm) since the curve is singular at the origin $r=0$.
The upper limit is set to $r_{{\rm max}}/a=11$ beyond which the amplitude of the NBS wave function is essentially zero.
Hence, the data for the pre-potential becomes meaningless beyond it.
Moreover, taking into account the cubic symmetry, we restrict our selves to $x,y,z\geq 0$.
The left panel of \Fig{ccbar_fit_ALL} shows the fit result where we see that most of the points lie close to the curve.
However, $\chi^2/N_{\rm d.o.f}$ is considerably large, being of the order of $\simeq1690$.
Let us refer to this fit  result as ``ALL'',  and label $\chi^2/N_{\rm d.o.f}$ for this as $\left[\chi^2/N_{\rm d.o.f}\right]_{\rm ALL}$ in order to distinguish it from other fits which will be presented in the following paragraphs.

One reason for the large $\chi^2/N_{\rm d.o.f}$  is the large deviation of the points near origin, namely those at $r/a=1,\sqrt{2},\sqrt 3$.
This is because they are computed by the discretized Laplacian containing the data at the origin which should make sense only in the continuum field theory.
Therefore, the data points at and near the origin deviate from the theoretical curve.
These points contribute significantly to the large value of $\chi^2/N_{\rm d.o.f}$, the sum being estimated to be $\simeq1460$.

The remaining $\chi^2/N_{\rm d.o.f}\simeq230$ may be attributed to the direction dependence due to RSB.
In order to see this point, we fit the spin-independent part for the following three representative directions: Series I ($\bm r = (n,0,0)$ ), Series \rmII ($\bm r = (n,n,0)$ ) and Series \rmIII ($\bm r = (n,n,n)$ ) with $n=0,1,2,3\cdots$.
There are residual directions with longer periods such as $(n,2n,0)$, but these contributions are less important.
In this case, the minimum of the fit range $r_{{\rm min}}$ needs to be determined carefully in order to fit the data adequately.
Otherwise, the fit yields unphysical results, e.g. negative Coulomb coefficients.
We set $r_{{\rm min}}$ such that $\chi^2/N_{{\rm d.o.f}}$ is reasonably small and the fit parameters become stationary.
These considerations lead us to the fit range $r_{{\rm min}}/a=4$ ($r_{{\rm min}}\simeq0.36$ fm).
The right panel of \Fig{ccbar_fit_ALL} shows the spin-independent part of the pre-potential for the three series using the Cornell+log function in this way.
Indeed, the fit yields reasonable chi-squares $\chi^2/N_{{\rm d.o.f}}\simeq1.5$, $1.5$ and $0.7$ for  Series I, Series \rmII and Series \rmIII, respectively.
We observe that Series I overestimates the Coulomb attraction near origin, thus separating out clearly from the other two towards the origin.

Now we evaluate the contribution of each series to $\left[\chi^2/N_{\rm d.o.f}\right]_{\rm ALL}$.
Let us define the chi-square for the three series by
\begin{equation}
\left[\chi^2/N_{\rm d.o.f}\right]_{\alpha}
\equiv
N_{\rm d.o.f}^{-1}\sum_i\left[\frac{(y^\alpha_i-f_{\rm ALL}(r))^2}{(\delta y^\alpha_i)^2}\right].
\end{equation}
Here, $y^\alpha_i$ denote the $i$-th data point in $\alpha$(= Series I, Series \rmII and Series \rmIII) direction
and $\delta y^\alpha_i$ denotes the statistical error of the data.
$\left[\chi^2/N_{\rm d.o.f}\right]_{\alpha}$ is approximated by
\begin{equation}
\left[\chi^2/N_{\rm d.o.f}\right]_{\rm \alpha}
\simeq
c_\alpha
\int _{r_0}^{11a} dr
\frac{(f_\alpha(r)-f_{\rm ALL}(r))^2}{(\delta f_\alpha(r))^2}
\end{equation}
where $f_\alpha(r)$ is given by $V_0^{\rm fit}(r)$ fit to Series $\alpha$.
The weight $c_\alpha$ is given by $c_\alpha=\frac{3}{a}, \frac{3}{\sqrt{2}a}$ and $\frac{1}{\sqrt{3}a}$ for  Series I, \rmII and \rmIII , respectively, considering the density of the data points along each direction.
We set the minimum of the integral $r_0=(1.5)a,(1.5)\sqrt2 a, (1.5)\sqrt 3a$ such that the data points at the tips of the first cubes ($r=a,\sqrt 2 a,\sqrt3 a$) are excluded from the integral.
Note that this approximation assumes each data point is rigorously on the fit curve.
On the other hand, the denominator $\delta f_\alpha(r)$ is the statistical error for each curve.
The error is expressed in terms of the errors of the parameters as
\begin{equation}
\delta f_\alpha(r)
\simeq
\left[
(\frac{\delta A}{r})^2 + (\delta B r)^2 + (\delta C \log(r))^2 + \delta v_0^2
\right]^{\frac{1}{2}}.
\end{equation}
The results are $\left[\chi^2/N_{\rm d.o.f}\right]_{\rm \alpha}\simeq 278$, $\simeq 1.2$ and $\simeq 0.17$, for Series I, Series \rmII and Series \rmIII, respectively.
The sum $\simeq 279$ is not far from the expected value 230.

Thus, we see that the large value of $\left[\chi^2/N_{{\rm d.o.f}}\right]_{\rm ALL}$ for ALL stems from two factors.
One is due to the singularity near the origin.
The other is due to the direction dependence caused by the RSB.
The approach to the continuum limit as $a\to 0$ may then be surmised as follows.
The data at the origin approaches negative infinity as $a$ becomes smaller.
Since the major error due to the discretized Laplacian is limited to the points ($r=a,\sqrt{2}a, \sqrt{3}a$), it tends to diminish in this limit.
Also, importantly,  the direction dependence is expected to become smaller by the power $\sim\mathcal O(a^2)$ as $a$ goes to 0. 
Thus, the fit curve for ALL and those for the three directions approach each other.
Overall,  the data points will fit to one well-defined curve in the continuum limit.

\begin{figure}[h]
\includegraphics[width=0.5\textwidth]{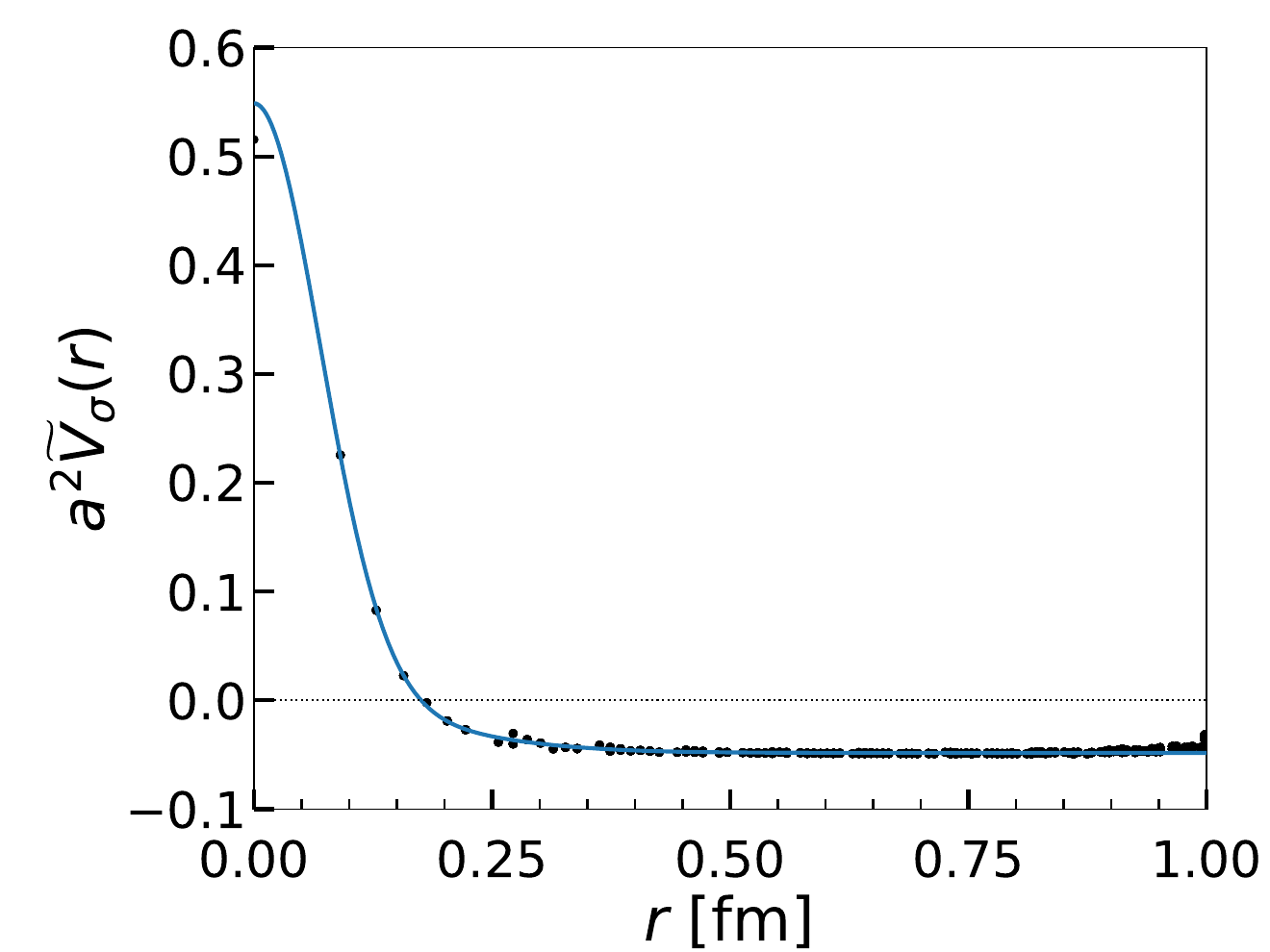}%
\caption{\label{ccbar_fit_2gauss}Spin-spin part of the pre-potential fit to 2-Gaussian function (solid line).
The black circles denote the pre-potential LQCD data.
The flat part of the pre-potential lies below 0 by the amount equal to $m_c\varDelta E_{hyp}$.}
\end{figure}

Next, we fit the spin-spin part of the pre-potential to the 2-Gaussian function in the range $1 \leq r/a \leq10$.
\Fig{ccbar_fit_2gauss} shows the fit result.
We see that most of the points lye on the curve, and $\chi^2/N_{{\rm d.o.f}}\simeq50.0$.
The fit result reproduces the NBS wave function reasonably well, as will be shown in the following subsection.
We refrain from separating the spin-spin part into the three directions because, otherwise, there are too few points that can be used for the fit analysis.
Fortunately, the symmetry breaking in the spin-pin part seems to be marginal.
This may be because the symmetry breaking in the PS state NBS wave function offsets that in the V state NBS wave function.

\subsection{Eigen value problem and the charm quark mass\label{subsect:ccbar_eigen}}
We put the fit result of the pre-potential $\widetilde{V}^{\rm fit}_{{\rm PS}}(r)=\widetilde{V}^{\rm fit}_{0}(r)-\frac{3}{4}\widetilde{V}^{{\rm fit}}_{\sigma}(r)$ so-obtained into the radial part Sch\"odinger equation \Eq{eq:P-wave-excitation}:
\begin{equation}
\left[-\frac{1}{r^2}\frac{d}{dr}\left(r^2\frac{d}{dr}\right)+\left\{\widetilde{V}^{\rm fit}_{{\rm PS}}(r)-\widetilde E_l\right\}+\frac{l(l+1)}{r^2}\right]\phi_l(r)=0.
\label{eq:Schrodinger_eq_radial}
\end{equation}
Note that  $\phi_0(r)=\phi_{{\rm PS}}(r)$, $\widetilde E_0 = 0$, $\phi_1(r)=\phi_{{\rm T}}(r)$ and $\widetilde E_1(r)=\widetilde E_{{\rm T}}$ follow from \Eq{eq:qqbar_pre-potentials}.
The Discretized Variable Representation  (DVR) method~\cite{DVR1,DVR2} is employed to numerically solve \Eq{eq:Schrodinger_eq_radial}.

\begin{figure*}[ht]
\includegraphics[width=\textwidth]{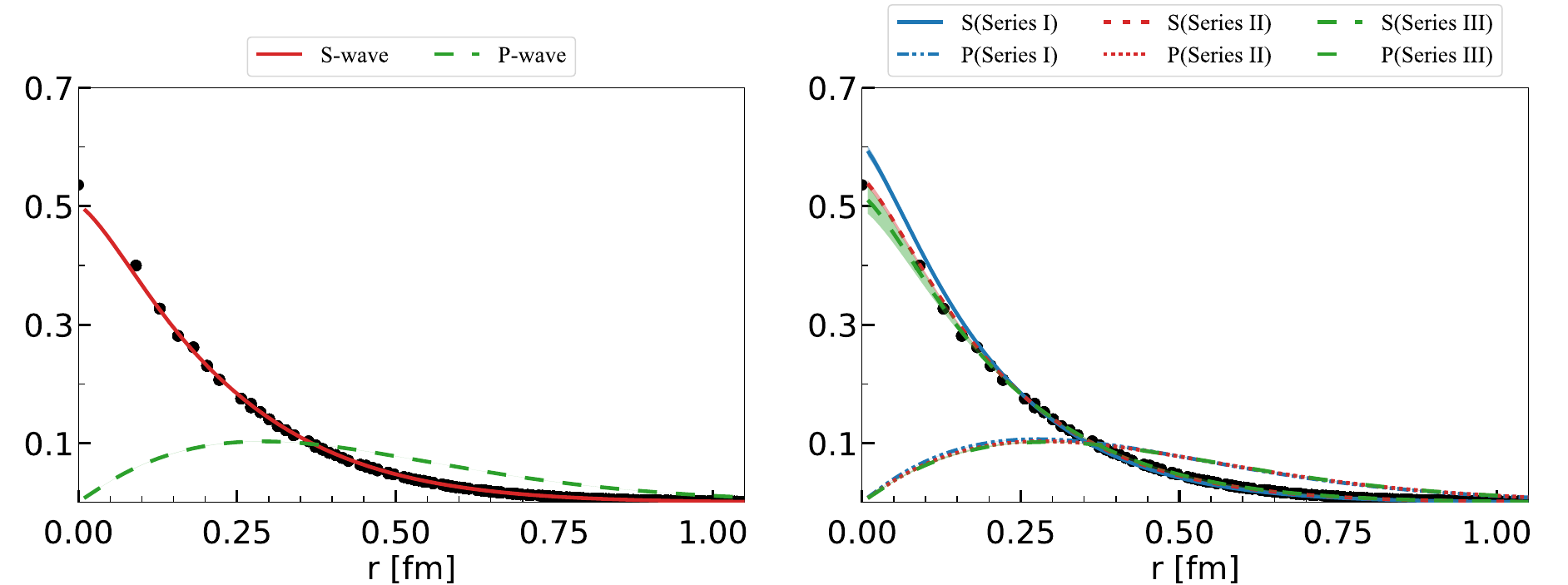}%
\caption{\label{ccbar_num}Numerical  solutions for S-wave (S) and P-wave (P) compared with the LQCD data (black dots).
 (left) from ALL and (right) from Series I, \rmII and \rmIII.
 Each shaded area denotes the statistical error.}
\end{figure*}

We solve the eigenvalue problem using the fit result from ALL.
The left panel of \Fig{ccbar_num} compares the numerical result for ALL and the NBS wave function LQCD data.
We see that the LQCD data lie close to the numerical solution except at the first two points near origin as expected from the discussions in the previous subsection.

We next use the fit from Series I, Series \rmII and Series \rmIII to estimate the direction dependence.
The right panel of \Fig{ccbar_num} compares the numerical solution for each series to the LQCD data.
The figure shows that the solutions differ in the short range region.
 According to ref.\cite{Ishizuka}, it is not surprising that Series I differs the most  from the S-wave
 \footnote{\label{ft:ishizuka}
 Ishizuka\cite{Ishizuka} argues that the NBS wave function in the $A_1$ representation of the cubic group equals S-wave up to angular momentum $l\ge 4$, hence it is of the form 
 $a v_0(r) + b v_4(r) {\cal Y}_{40}(\theta,\phi)+\cdots$ where $a$ and $b$ are constants. Thus, the wave function deviates from the S-wave most on the principal axes 
 $X$, $Y$ and $Z$ of the cube if the residue is negligible. In the present context, this means that Series I is expected to differ most from the S-wave.}.
On the other hand, we cannot determine definitively or convincingly which is the better, Series \rmII or Series \rmIII within the bounds of numerical precision. 
For instance, consider the residual sum of squares (RSS) $\delta_{\rm RSS}\equiv \sum_{r}\left\{\phi_{\rm num}(r)-\phi_{\rm NBS}(r)\right\}^2$ between the S-wave numerical solution and the NBS wave function.
Then, $\sqrt{\delta_{\rm RSS}}\simeq8\cdot10^{-2}$,  $2\cdot10^{-2}$, $3\cdot10^{-2}$ and $5\cdot10^{-2}$ for Series I, Series \rmII, Series \rmIII and ALL, respectively. 
The last three are of the same order of the numerical precision of the normalization $\simeq10^{-2}$.
See footnote.\ref{ft:ishizuka} for the marked difference of Series I.
The P-wave, on the other hand, is weakly affected by the direction dependence near origin, and thus the solutions for the series coincide.

\begin{figure}
\includegraphics[width=0.5\textwidth]{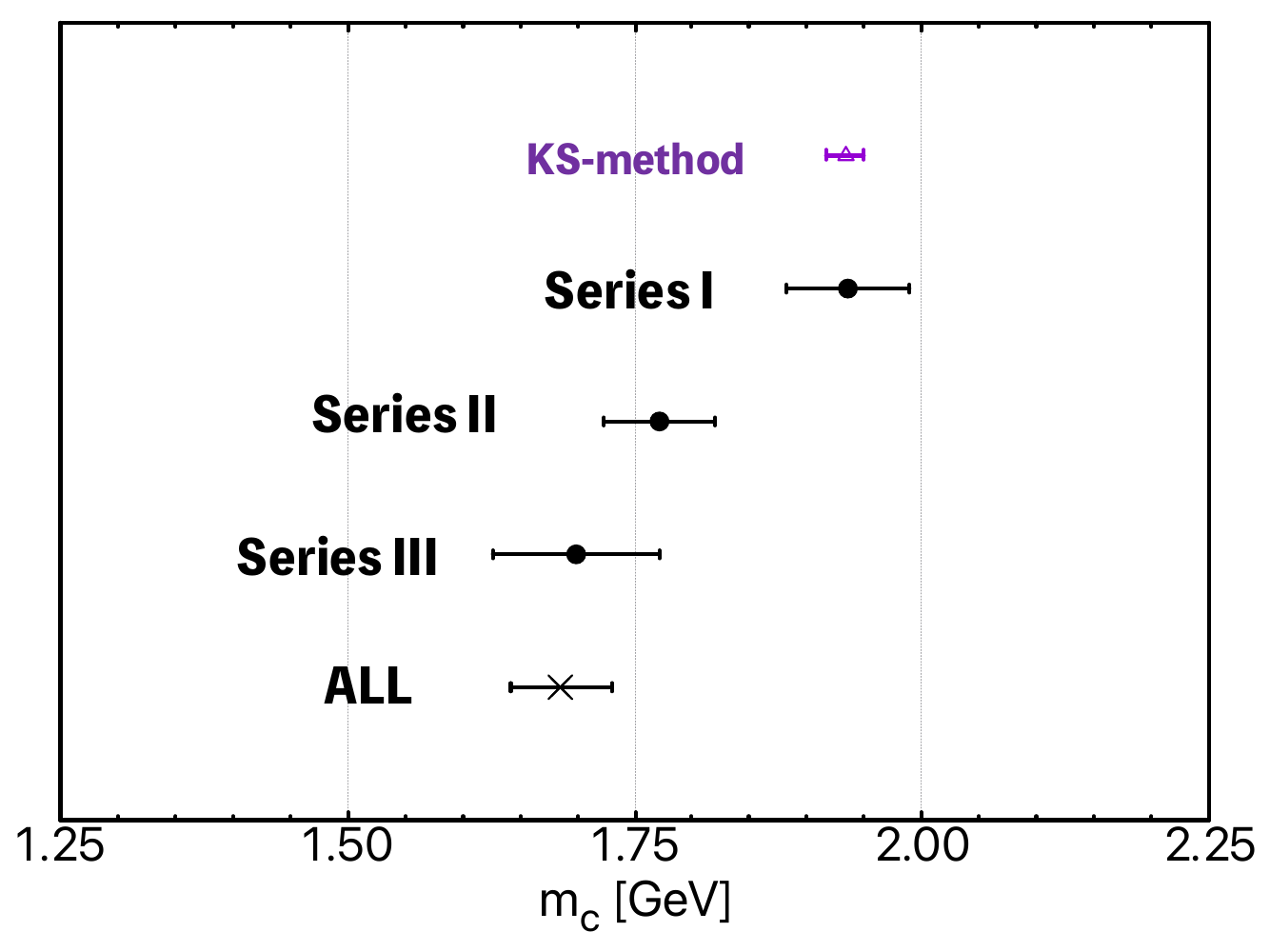}%
\caption{\label{KS_vs_WI}Charm quark mass from ALL and SeriesI, \rmII, \rmIII compared to that of the Kawanai-Sasaki method.}
\end{figure}

Using the excitation pre-energy $\tilde E_{{\rm PS,T}}$, the charm quark mass is given by $m_c=\frac{\tilde E_{{\rm PS,T}}}{M_{{\rm T}}-M_{{\rm PS}}}$.
 \Table{charm_quarm_mass} summarizes $m_c$ where we see that each mass lies between $1.9$ GeV and $1.7$ GeV.
The difference of $\sim200$ MeV is the systematic error due to the RSB.
\Fig{KS_vs_WI} compares the charm quark mass for ALL, Series I, Series \rmII and Series \rmIII with that obtained from the Kawanai-Sasaki method $m_c=1.933(17)$ GeV.
In the figure, we see that our result and the value from the Kawanai-Sasaki method roughly agree.
The slight difference comes from the fact that our method uses the P-wave excitation energy while the Kawanai-Sasaki method uses the hyperfine splitting.
Following the convention, we use the charm quark mass obtained from ALL hereafter for necessary conversions since there is no definitive choice among the three directions within the bounds of numerical precision.

\begin{table}
		\caption{\label{charm_quarm_mass}Summary of the charm quark mass.}
	\begin{center}
		\begin{tabular}{c|cccc}
		\hline
  	  	directions & ALL & Series I & Series \rmII & Series \rmIII \\
		\hline 
		$m_c$ [GeV] & 1.686 (44) & 1.936 (53) & 1.771 (48) &  1.699 (73)\\
		\hline
   	 	\end{tabular}
		\label{table:charm_mass}
	\end{center}
\end{table}

Using the charm quark mass $m_c=1.686 (44)$ GeV and the pre-potential, the potential is given by $V_0(r) = \frac{1}{m_c}\tilde V_0(r)+\overline{E}_{1S}$.
\Table{ccbar_pot_prams} summarizes the result of the coefficients of the Cornell+log function $F(r)=-\frac{A}{r}+B r +C\log(r/a) +v_0$ fit to ALL and Series I, \rmII and \rmIII.
The Coulomb coefficient $A$ depends strongly on the direction because it is determined in the short range region as noted before.
On the other hand, the coefficients $B$, $C$ and $v_0$ have smaller direction dependence since they are determined from the long range region.

The spin-spin potential is given by $V_\sigma(r)=\frac{1}{m_c}\tilde{V}_\sigma(r)+\varDelta E_{\rm hyp}$.
The parameters of the 2-Gauss function \Eq{eq:2Gauss} fit to the spin-spin potential are $A_1=0.135(4)$ GeV, $B_1=19.34(38)$ fm$^{-2}$, $A_2=1.540(41)$ GeV, $B_2=104.7(4)$ fm$^{-2}$ and $v_\sigma=-0.0170(38)$ GeV.
 
\begin{table}
		\caption{\label{ccbar_pot_prams}Summary of the parameters for the Cornell+log function fit to the spin-independent part of the $c\bar c$ potential.}
	\begin{center}
		\begin{tabular}{c|c c c c }
		\hline
  	  	direction & $A$ [GeV$\cdot$fm]  & $B$ [GeV/fm]         & $C$ [GeV]           & $v_0$ [GeV]\\
		\hline 
		ALL              &0.080(2)&0.748(26)&0.328(14)&-0.964(70)\\
		Series I         &0.118(4)&0.647(28)&0.331(11)&-0.856(74)\\
		Series \rmII   &0.098(4)&0.695(37)&0.312(11)&-0.867(74)\\
		Series \rmIII  &0.089(9)&0.737(60)&0.300(16)&-0.880(75)\\
		\hline
   	 	\end{tabular}
	\end{center}
\end{table}

\section{\label{cD_Num}Numerical results for $cD$ system}
\subsection{\label{cD_NBS_lev} $\Lambda_c$ mass}
Let us start with the $\Lambda_c(\frac12)$ baryon correlator $C^{\frac{1}{2}}_{\alpha\beta}(t)$ given by
\begin{equation}
  C^{\frac{1}{2}}_{\alpha\beta}(t)
  \equiv
  \frac1{V}
  \sum_{\bm x}
  \left\langle 0 \left|
  T\left[ \mathcal B_\alpha(\bm x,t_{{\rm sink}}) \overline{\mathcal B}_\beta(t_{{\rm src}}) \right]
  \right| 0 \right\rangle,
  \label{eq:corr_Lambda_c}
\end{equation}
where $\mathcal B_\alpha=\left[u^T C\gamma_5 d\right]c_{\alpha}$ such that $\mathcal B_\alpha(\bm x,t_{{\rm sink}})$ denotes the $\Lambda_c(\frac12)$ point sink operator and $\overline{\mathcal B}_\alpha(t_{{\rm src}})$ denotes the wall source operator.
The operator $\mathcal B_\alpha(\bm x,t)$ couples to both positive parity and negative parity states.
Therefore, the correlator has components corresponding to the propagation between the opposite parity states as well as between the same parity states.
To eliminate the propagation between the opposite parity states, we act the projection operator $P^\pm_{\alpha\beta}=\left[\frac{\bm 1 \pm \gamma_0}2\right]_{\alpha\beta}$ to $C^{\frac{1}{2}}_{\alpha\beta}(t)$ as
\begin{equation}
  C^{\frac{1}{2}^\pm}(t)
  =
  \sum_{\alpha,\beta}
  P^\pm_{\alpha\beta}C^{\frac{1}{2}}_{\alpha\beta}(t),
\end{equation}
where $C^{\frac{1}{2}^+}(t)$ is the correlator of the positive-to-positive propagation and $C^{\frac{1}{2}^-}(t)$ that of the negative-to-negative parity propagation.
On the other hand, the $\Lambda_c(\frac{3}{2})$ baryon two-point correlator is given by
\begin{equation}
  C^{\frac{3}{2}^\pm}(t)
  =
  \sum_{\alpha,\beta}
      P^\pm_{\alpha\beta}
  \sum_{i,j}
          \sum_{\alpha'}
  C_{\alpha'\beta;ij}(t)
    P^{ji}_{\frac{3}{2}:\alpha\alpha'},
  \label{eq:RS_corr}
\end{equation}
where $C_{\alpha\beta;ij}(t)$ is defined as 
\begin{equation}
\begin{split}
  C_{\alpha\beta;ij}(t)
  \equiv
  \frac1{V}
  \sum_{\bm x}
  \left\langle 0 \left|
  T\left[ \mathcal R_{\alpha,i}(\bm x,t_{{\rm sink}}) \overline{\mathcal R}_{\beta,j}(t_{{\rm src}}) \right]
  \right| 0 \right\rangle.
  \end{split}
\end{equation}
Here, with $\mathcal R_{\alpha,i}=\left[u^TC\gamma_5\gamma_i d\right](\gamma_{5})_{\alpha\alpha'}c_{\alpha'}$ being the Rarita-Schwinger vector-spinor operator~\cite{Leinweber}, $ \mathcal R_{\alpha,i}(\bm x,t_{{\rm sink}})$ is the sink operator and $\overline{\mathcal R}_{\alpha,i}(t_{\rm src})$ is the wall source operator.
The labels $i$ and $j$ denote the vector indices.
The operator ${P}^{ij}_{\frac{3}{2}:\alpha\beta}= \delta^{ij}\bm 1_{\alpha\beta}-\frac{1}{3}\left(\gamma^i\gamma^j\right)_{\alpha\beta}$ in \Eq{eq:RS_corr}
represents the projection to the spin $3/2$ state in the rest frame~\cite{benmerrouche} whereby the spin 1/2 components are eliminated from $C_{\alpha\beta;ij}(t)$~\cite{Leinweber,alexandrou_RS,zanotti,bahtiyar}.
Note that the statistical noises of the correlators are reduced by making use of an identity $ C^{J^\pm}(t)=-C^{J^\mp}(T-t)$.

\begin{figure}[h]
\includegraphics[width=0.48\textwidth]{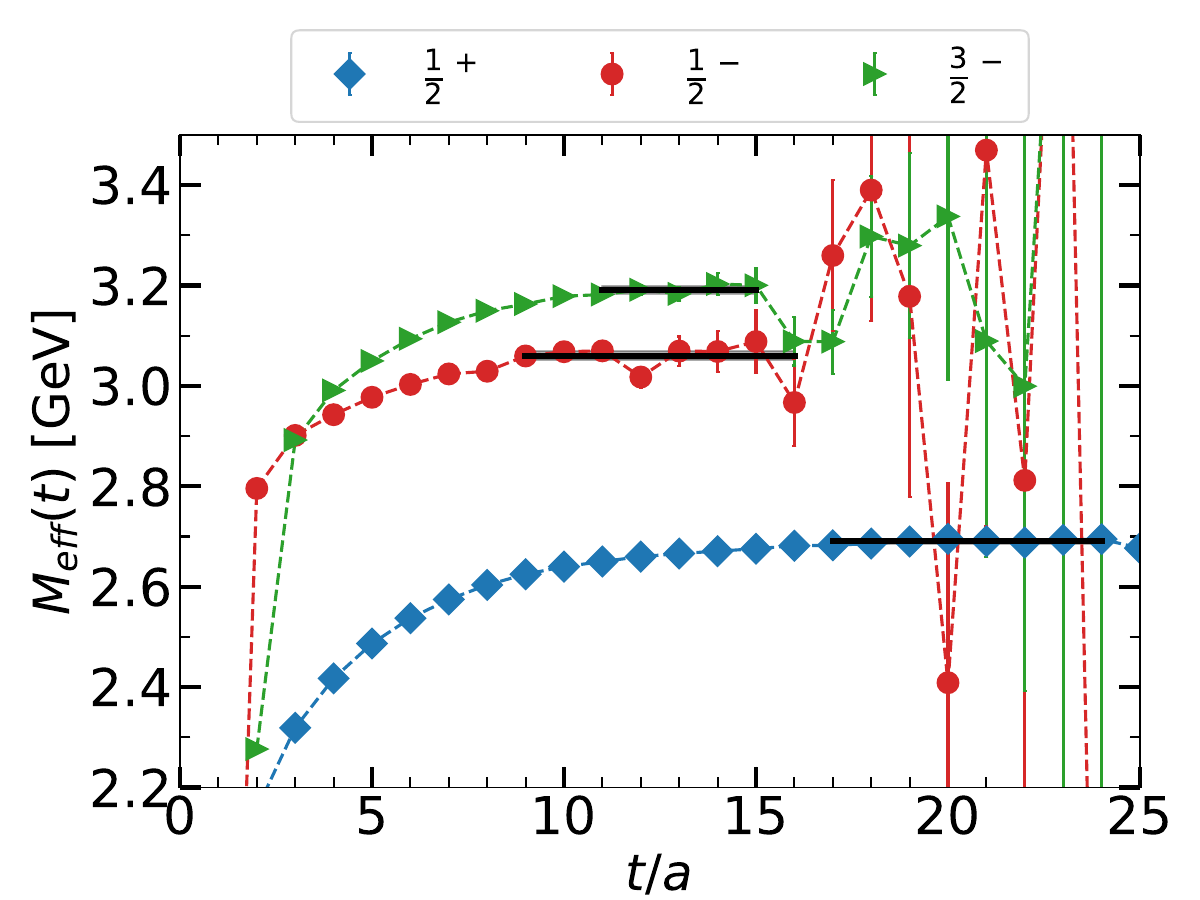}%
\caption{Effective mass plots for $\Lambda_c\left(\frac{1}{2}^+\right)$ (blue diamond), $\Lambda_c\left(\frac{1}{2}^-\right)$ (red circle) and $\Lambda_c\left(\frac{3}{2}^-\right)$ (green triangle).  \label{Lambda_c_EFM}}
\end{figure}

 The effective mass of the baryon is given by 
$M_{\rm eff}(t)=\log\left(\frac{C^{J^P}(t)}{C^{J^P}(t+1)}\right)$.
The baryon mass is extracted by fitting $A\exp(-Mt)$ to the correlator in the plateau region.
\Fig{Lambda_c_EFM} shows the effective mass plot and the baryon mass extracted by the fitting analysis.
The fitting is adequately carried out as shown in the figure with the values of the chi-square $\chi^2/N_{{\rm d.o.f}}< 1.0$ for all the states of our interest.
 \Table{tab:baryon_mass_table} summarizes the values of the baryon mass.
 We obtain the LS average $\overline M_p = \frac{1}{3}\left(2M_{(3^-/2)}+M_{(1^-/2)}\right)=1.447(3)$ GeV and the P-wave excitation energy $\varDelta E_{s,p}=0.457(7)$ GeV using the $\Lambda_c$ masses.
 
The same analysis is applied to the $\Sigma_c(\frac{1}{2}^+)$ baryon and the nucleon.
The mass values are $M_{\Sigma_c}=2.794 (3)$ GeV and $M_N=1.574(18)$  GeV.
All the $\Lambda_c$ baryon states are bellow $\Sigma_c+\pi$ threshold and are thus bound.

\begin{table}[h]
\caption{\label{tab:baryon_mass_table}Values of the $\Lambda_c$ baryon masses. The last row shows the fit range.}
		\begin{tabular}{ c c c } 
		\hline\hline
		     &  Mass [GeV] &  fit range \\ \hline
		$\Lambda_c(\frac{1}{2}^+)$ &  2.691 (5) & $17 \leq t/a \leq 24$ \\
		$\Lambda_c(\frac{1}{2}^-)$ &   3.060 (9) & $9 \leq t/a \leq 16$\\
		$\Lambda_c(\frac{3}{2}^-)$ &   3.192 (8)& $11 \leq t/a \leq 15$ \\
		\hline\hline
		\end{tabular}
\end{table}

\subsection{\label{cD_NBS_pot} $cD$ NBS wave function and the pre-potential}
In order to extract the NBS wave function, we begin with the definition of the $cD$ four-point correlator $G_{cD,\alpha\beta}(\bm r, t)$ given by
\begin{eqnarray}
  \lefteqn{
    G_{cD,\alpha\beta}(\bm r, t)
  }
  \\\nonumber
  &\equiv&
  \frac1{V}
  \sum_{\bm \Delta}
  \left\langle 0 \left|
  T\left[
    D_a(\bm r + \bm \Delta, t_{{\rm sink}})
    c_{a,\alpha}(\bm \Delta, t_{{\rm sink}})
    \cdot
\overline{\mathcal B}_\beta(t_{{\rm src}})
    \right]
  \right| 0 \right\rangle,
\end{eqnarray}
where $a$ denotes the color index.
We  focus on the degenerated components $(\alpha,\beta)=(1,1),(2,2)$ which represent the positive-to-positive parity propagation, and omit the spinor indices hereafter.
The degenerated components are later averaged over to reduce the statistical error.

As demonstrated for the $c\bar c$ four-point correlator in subsection\ref{Numerical_results}, the $cD$ four-point correlator is spectrally decomposed as
\begin{equation}
G_{cD}(\bm r, t)
=
\sum_n
\psi^{(n)}_{cD}(\bm r)
a_{n}
e^{-M_{n} t}
\end{equation}
where $\psi^{(n)}_{cD}(\bm r)$ , $a_{n}=\langle n|\overline{\mathcal B}(0)|0\rangle$, and $M_{n}$ denote the NBS wave function, the overlap, and the mass of the $n$-th excited state, respectively.
In the large time region, the excited states are suppressed and the ground state $\Lambda_c(\frac12^+)$ becomes dominant.
For this reason, we denote $\psi^{(0)}_{cD}(\bm r, t)$ as $\psi_{\Lambda_c}(\bm r, t)$.
To proceed, we project the wave function to the S-wave by the $A_1$ projection \Eq{eq:A1+}.

\begin{figure}[h]
\includegraphics[width=0.5\textwidth]{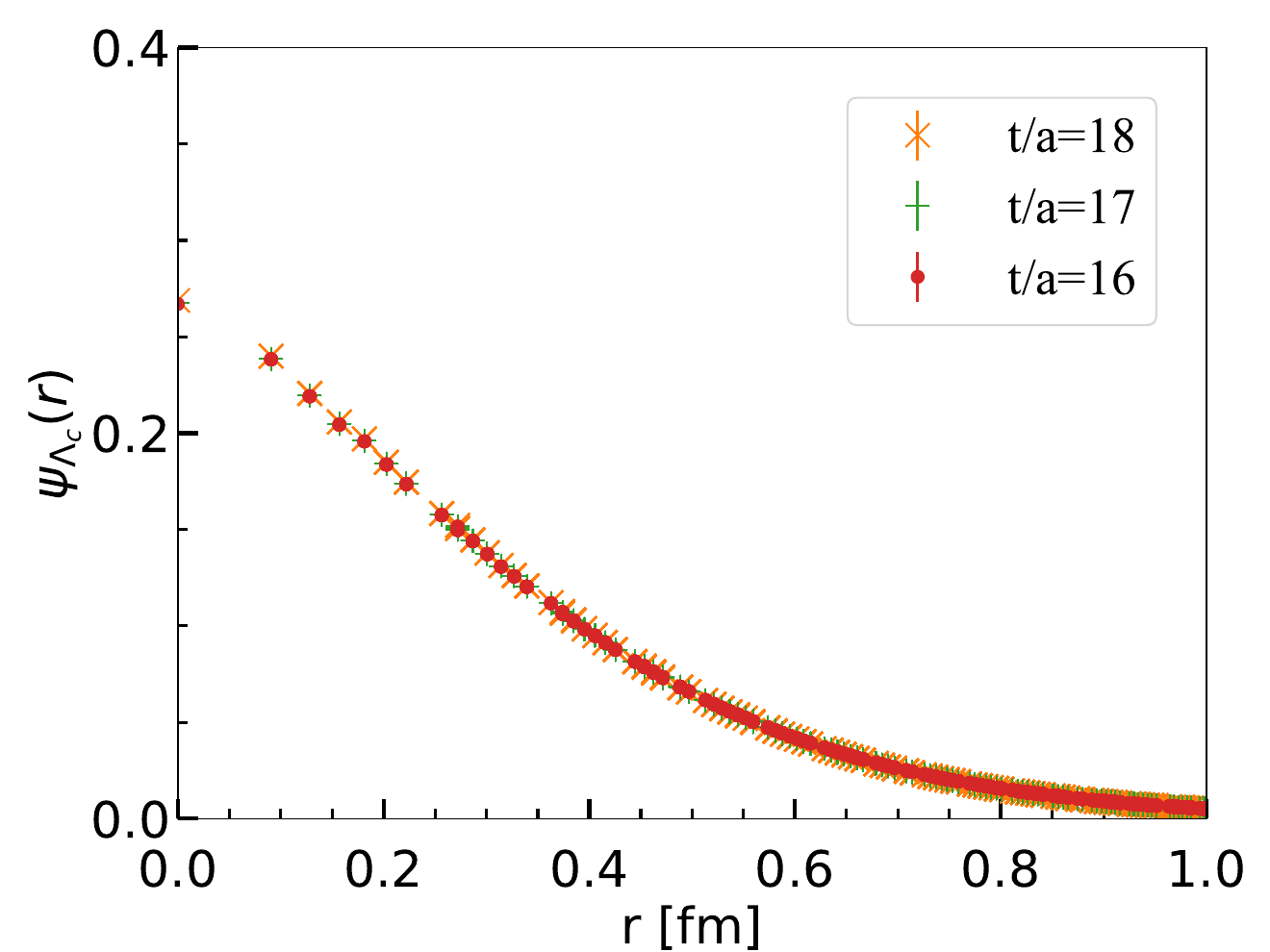}%
\caption{$cD$ NBS functions for the $\Lambda_c(\frac{1}{2}^+)$ state at $16 \leq t/a \leq 18$. 
Each wave function is normalized as $\sum_{r \in V}\bm{r}^2\psi_{\Lambda_c}^2(\bm r) = 1$.}
\label{cD_WF}
\end{figure}

\Fig{cD_WF} shows the $cD$ NBS wave functions in several representative time slices.
As in the case of $c\bar c$, we focus on the wave function at $t=17$ here as the representative one from now on.

Note that the $cD$ NBS wave function is spatially extensive and experiences reflection at the boundary.
Then, the wave function is deformed from the S-wave especially in the long range region~\cite{Ishizuka}.
Our observation is corroborated by the growing systematic uncertainty at $r/a\simeq 10$ ($r\sim0.91$ fm) due to the reflection.
In order to avoid this undesirable effect, we limit the range to $r/a\leq 9$  ($r\lessapprox 0.82$ fm).

\subsection{Eigen value problem and the Diquark mass}
\begin{figure*}[ht]
\includegraphics[width=\textwidth]{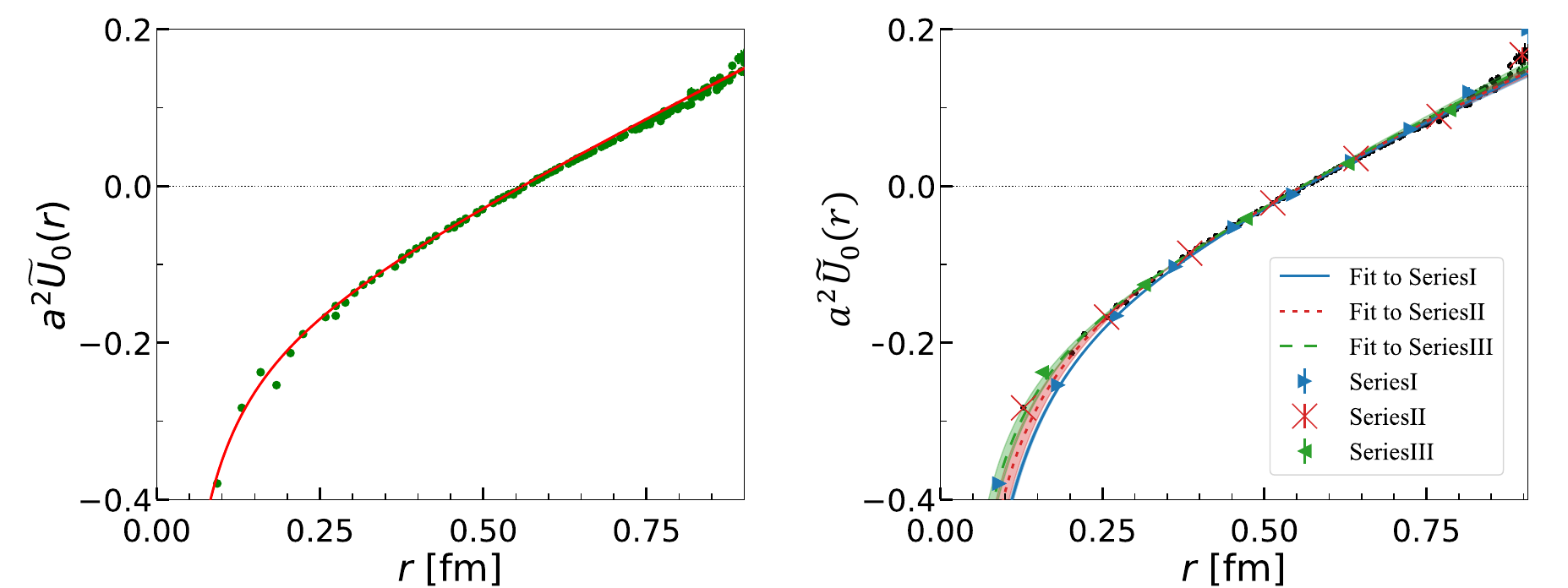}%
\caption{\label{cud_fit_ALL}Cornell function fit to the $cD$ pre-potential (black dots).
Left panel shows the fit to ALL and right shows the fit to Series I,\rmII and \rmIII. }
\end{figure*}

We construct the $cD$ pre-potential in the same way as for the $c\bar c$ pre-potential in \Sect{Meson_Num}.
Then,  we first fit to ALL in the range $1 \leq r/a \leq 9$ ($0.1\lessapprox r \lessapprox 0.82$ fm) to the Cornell function $f(r)=-A/r+Br+C$ without the log term.
As the left panel of \Fig{cud_fit_ALL} shows, most of the data points lie close to the curve.
Thus this fitting is adequate.
Note that the fit yields relatively large chi-square of $\chi^2/N_{{\rm d.o.f}}\simeq23$ as in the case of $c\bar c$ system due to the singularity near origin and the direction dependence.
Next, we fit the data points for the three series separately as before by setting the fit range appropriately to $3 \leq r/a \leq 9$ ($0.27\lessapprox r \lessapprox 0.82$ fm).
The fit is adequate as shown in the right panel of \Fig{cud_fit_ALL}, yielding $\chi^2/N_{{\rm d.o.f}}$; $\chi^2/N_{{\rm d.o.f}}\simeq2.7$, $0.47$ for $0.24$ for Series I, \rmII and \rmIII, respectively.

\begin{figure*}[ht]
\includegraphics[width=\textwidth]{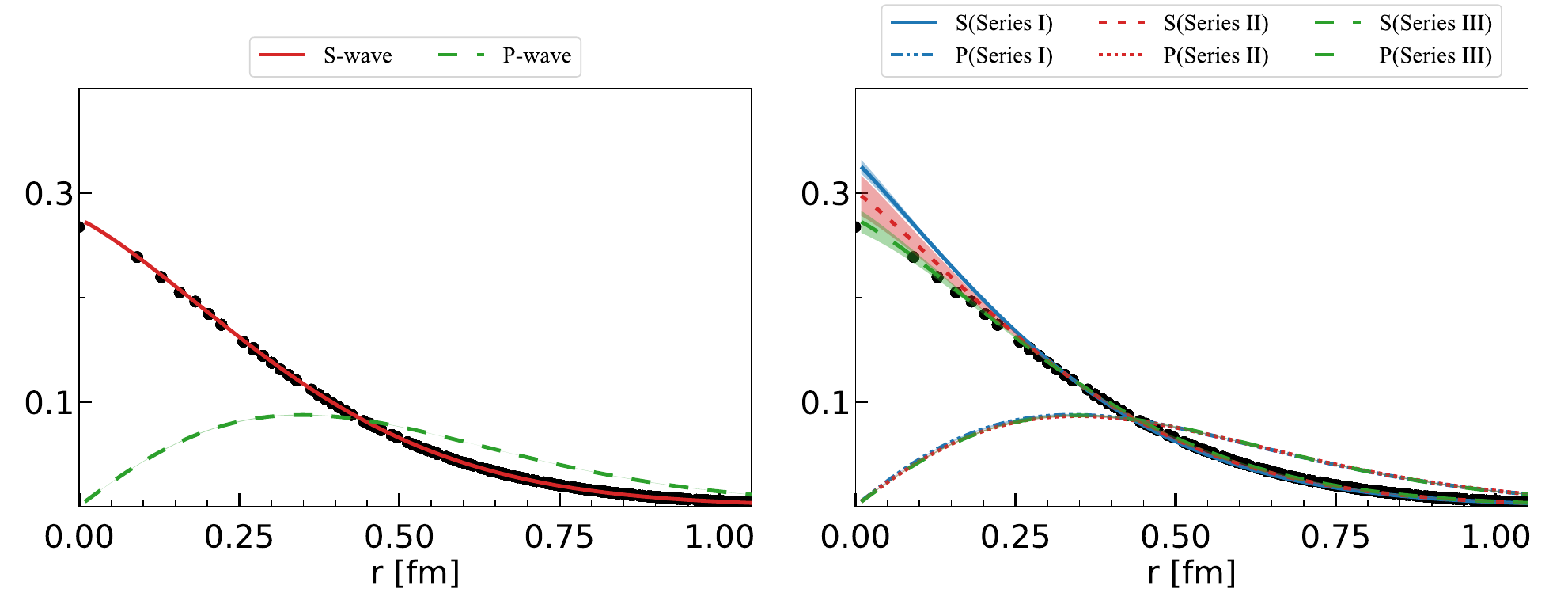}%
\caption{\label{ALL_cD_num}Numerical solution from (left) ALL and (right) Series I, \rmII and \rmIII.
The LQCD data of the NBS wave function (black circles) is shown for comparison.}
\end{figure*}

Substituting the fit result into \Eq{eq:Schrodinger_eq_radial}, we solve the eigenvalue problem for the S-wave and the P-wave.
The left panel of \Fig{ALL_cD_num} shows the solution for ALL whereas the right panel shows the numerical solutions of the three series.
In contrast to the $c\bar c$ system, the fit to ALL happens to reproduce the NBS wave function very well.
We see in the right panel of \Fig{ALL_cD_num} that the numerical result for the series and the LQCD data roughly agree.
The wave functions of ALL and \rmIII almost coincide.
As was discussed in subsection.\ref{subsect:ccbar_eigen} in the context of the spherical harmonic expansion, Series I is known to be the farthest from the S-wave.
Series \rmIII reproduces the NBS wave function seemingly the best.
The values of the RSS are $\sqrt{\delta_{\rm RSS}}\simeq1\cdot10^{-2}$, $8\cdot10^{-2}$, $4\cdot10^{-2}$ and $9\cdot10^{-3}$ for ALL, Series I, Series \rmII and Series \rmIII, respectively.
The P-wave, on the other hand, is weakly affected by the direction dependence near origin, and thus the solutions for the series are indistinguishable.

\begin{figure}
\includegraphics[width=0.5\textwidth]{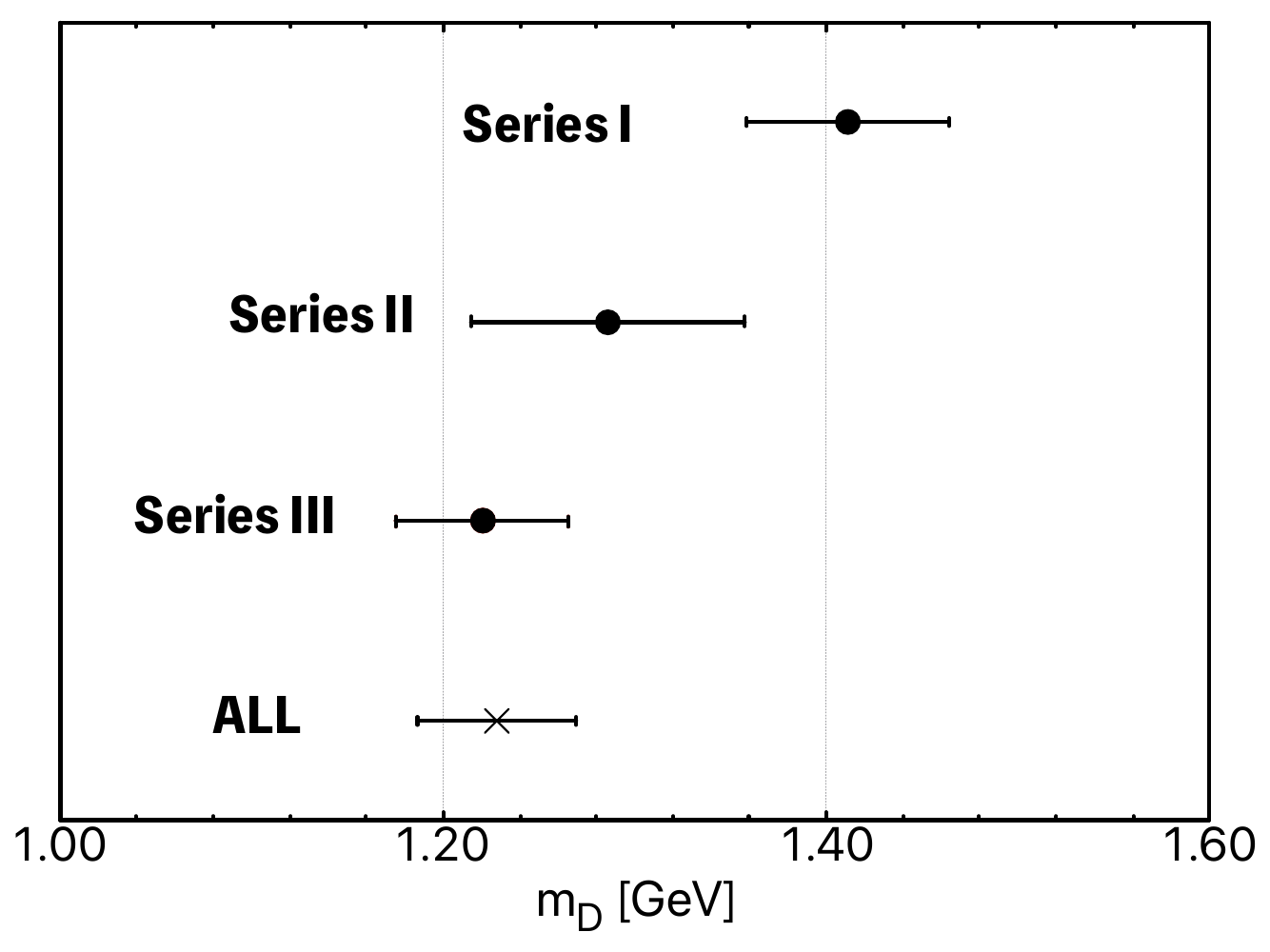}%
\caption{\label{direction_diquark}Direction dependence of the diquark mass.}
\end{figure}

The reduced mass $\mu_{cD}$ is obtained by \Eq{eq:mu_cd} using $\tilde E_{\rm PW}$ and the $\Lambda_c$ mass obtained in subsection \ref{cD_NBS_lev}.
Then, the diquark mass is obtained by substituting the charm quark mass $m_c$ into \Eq{eq:m_d}.
\Table{diquark_mass} summarizes the diquark mass obtained from each series.
The value of the diquark mass is roughly consistent with naive expectations from the constituent quark picture, i.e. the $\rho$ meson mass $M_\rho=1.098(5)$ GeV and twice the constituent mass $\frac{2}{3}M_N=1.049(12)$ GeV.
We see in the table that the diquark mass obtained from ALL, Series \rmII and Series \rmIII are close to each other while that from Series I is larger than the others.
This becomes more evident when we plot the mass as shown in \Fig{direction_diquark}.
For the sake of consistency with the $c\bar c$ system, let us use the diquark mass from ALL hereafter for necessary conversions.

\begin{table}[h]
		\caption{\label{diquark_mass}Summary of the diquark quark mass.}
	\begin{center}
		\begin{tabular}{c|cccc}
		\hline
  	  	directions & ALL & Series I & Series \rmII & Series \rmIII \\
		\hline 
		$m_D$ [GeV] & 1.273 (44) & 1.470 (57) & 1.335 (77) &  1.264 (49)\\
		\hline
   	 	\end{tabular}
		\label{table:charm_mass}
	\end{center}
\end{table}

Now, the $cD$ potential is given by $U_0(r) = \frac{1}{2\mu_{cD}}\tilde U_0(r)+E_{\Lambda_c(\frac12^+)}$.
\Table{cD_pot_prams} shows the coefficients of the Cornell function fit to the $cD$ potential.
We see that the direction dependence of the Coulomb coefficient is large compared to that of the string tension.
This is because the Coulomb term is determined in the short range region where the discretization errors are large.
It should also be noted that the string tension has some direction dependence as \Table{cD_pot_prams} shows.
One reason for this is that the string tension is affected by the direction dependence of the Coulomb coefficient because the region where the linear part is  overwhelmingly dominant is not reached.
Moreover, the reflected waves cause systematic errors to the linear part of the $cD$ potential, making it difficult to reach the desired asymptotic region.
It is necessary to calculate with a larger lattice as well as to obtain information from excited state wave functions with greater spatial extents.
In this way, the string tension may be determined with higher precision.

\begin{table}[h]
		\caption{\label{cD_pot_prams}Summary of the Cornell function fit to the $cD$ potential.}
	\begin{center}
		\begin{tabular}{c|c c c }
		\hline
  	  	direction & $A$ [GeV$\cdot$fm]  & $B$ [GeV/fm] & $C$ [GeV]\\
		\hline 
		ALL               &0.065(2)&1.315(24)&-0.889(34)\\
		Series I          &0.107(7)&1.157(53)&-0.728(52)\\
		Series \rmII   &0.089(16)&1.195(78)&-0.778(82)\\
		Series \rmIII &0.066(11)&1.300(71)&-0.876(68)\\
		\hline
   	 	\end{tabular}
	\end{center}
\end{table}

\section{\label{ccbar_vs_cD}Comparing $cD$ potential and $c\bar c$ potential}

\begin{figure*}[ht]
\includegraphics[width=\textwidth]{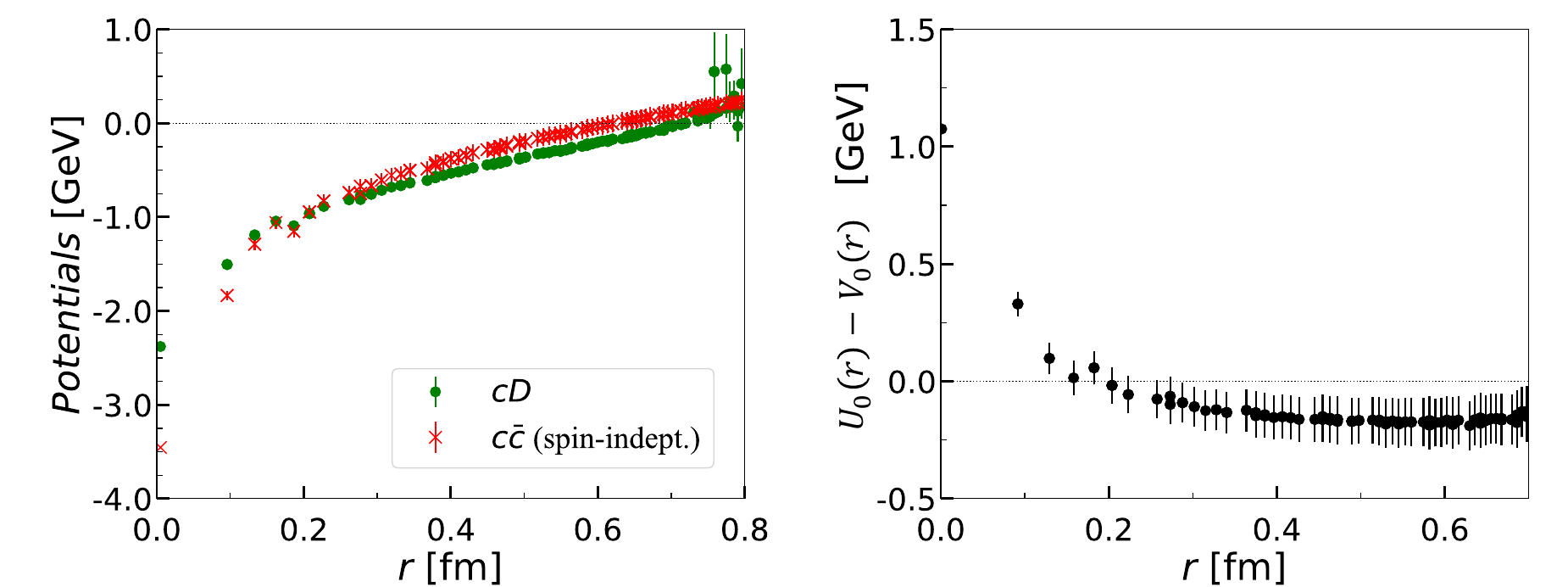}%
\caption{\label{cud_vs_ccbar} (left) $cD$ potential $U_0(r)$ compared with the spin-independent part of the $c\bar c$ potential $V_0(r)$. 
(right) Difference between $U_0(r)$ and $V_0(r)$.}
\end{figure*}

\Fig{cud_vs_ccbar} compares the $cD$ potential $U_0(r)$ and the spin-independent part of the $c\bar c$ potential $V_0(r)$.
It shows that their data points are almost parallel in the region $r\geq 0.4$ fm.
The difference between the two sets is essentially a constant as shown on a magnified scale in the right panel of \Fig{cud_vs_ccbar}.
Thus the linearly rising part of $U_0(r)$ and $V_0(r)$ cancel each other.
This behavior can be understood as follows.
First, the anti-charm quark and the diquark are in the same color representation $\overline{\bm 3}$.
Thus the $c\bar c$ potential and the $cD$ potential are identical in principle if it were not for the internal structures of the diquark.
Though the diquark is not point-like~\cite{diquark_size}, diquark structure is less important at long distances and the color representation governs the potential.
Indeed, it is reported in the earlier LQCD works~\cite{TTTakahashi,Suganuma-QQQ,TTTakahashi-Lett} that the diquark limit of the static three quark potential (QQQ potential) and the static $Q\bar Q$ potential coincide. 
  
At short distances,  on the other hand, the magnitude of the Coulomb attraction of the $cD$ potential is smaller than that of the  $c\bar c$ potential.
This mitigation of the Coulomb attraction is most likely caused by the internal structure of the diquark, namely its spatial extent.
In fact, the refrences~\cite{jido,Jido-size} shows that the Coulomb attraction is suppressed when the diquark has a finite spatial extent.
Note that the diquark limit in the QQQ system realizes a point-like diquark. 
Thus, the Coulomb attraction of the QQQ potential in the diquark limit is not weakened by the finite size effect of the diquark. 
Hence, it stands to reason that the attraction of our $cD$ potential is different from the diquark limit of the QQQ potential.

\begin{figure}
\includegraphics[width=0.5\textwidth]{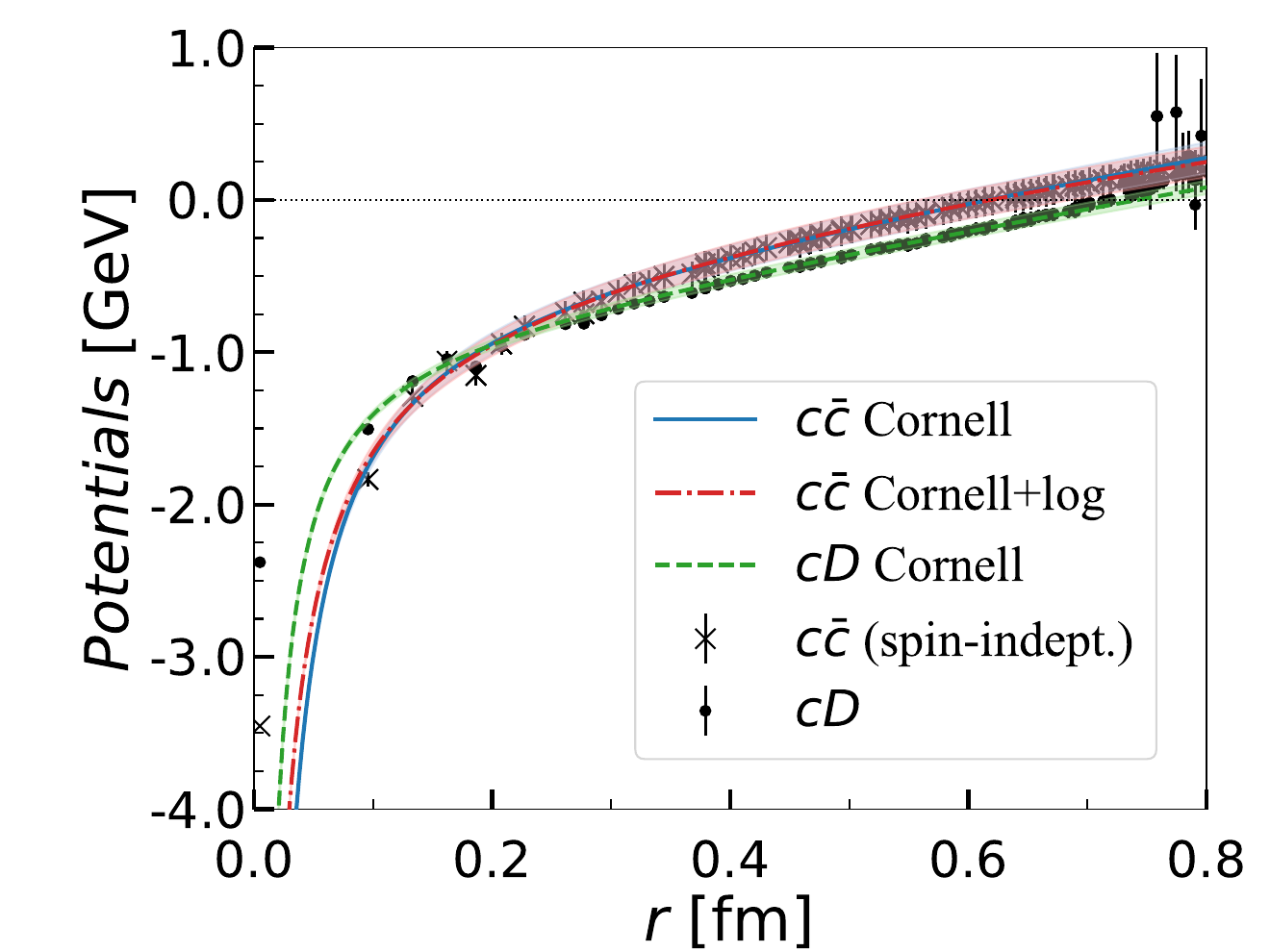}%
\caption{\label{cud_vs_ccbar_fit} Cornell function fit to the $cD$ potential and to the spin-independent part of the $c\bar c$ potential. 
	We also show the Cornell+log type function fit to the $c\bar c$ potential for a reference.}
\end{figure}

\Fig{cud_vs_ccbar_fit} compares the Cornell function fit to $U_0(r)$ and $V_0(r)$.
\Table{ccbar_cud_pot} summarizes the fitting parameters where we immediately see that the values of the string tension are roughly the same as expected from \Fig{cud_vs_ccbar}.
\begin{table*}
		\caption{\label{ccbar_cud_pot} Comparison of the the $cD$ potential and the $c\bar c$ potential fit to the Cornell function for ALL in the range $1\leq r/a \leq 9$.}
	\begin{center}
		\begin{tabular}{c| c c c}
  	  	  \hline 
		  & $A$ [GeV$\cdot$fm]& $B$ [GeV/fm] & C [GeV] \\
		\hline 
		 $cD$ &0.065(2)&1.315(24)&-0.889(34)\\
		$c\bar c$ &0.124(3)&1.247(24)&-0.576(80)\\
		\hline
   	 	\end{tabular}
	\end{center}
\end{table*}

\section{\label{Ketsuron}Conclusions}
We have reported on the novel and consistent method which we developed to evaluate the charm-diquark interaction potential, the diquark mass and the charm quark mass.
Specifically, we applied our method by considering the $\Lambda_c$ baryon as a bound state of a charm quark and a scalar-diquark. 
The diquark mass obtained from our method was $m_D=1.273 (44)$ GeV which does not contradict the conventional estimates; our diquark mass is slightly above the $\rho$ meson mass $m_\rho=1.098(5)$ GeV and twice the constituent quark mass $2m_N/3=1.049(12)$ GeV.
When applied to the charmonium, our method yielded $m_c=1.686 (44)$ GeV for the charm quark mass.
This is $\sim 250$ MeV smaller than the value $m_c=1.933(17) $ GeV obtained by the Kawanai-Sasaki method.
Our charm-diquark potential is given by the well-known Cornell potential where the Coulomb attraction was found considerably weaker than that of the $c\bar c$ potential. 
The weakening is most likely due to the structure of the diquark, namely its spatial extent.
These quantities allow us to construct a QCD-based quark-diquark model that can be used to investigate the hadron levels and structures.

In this work, we observed that there are three major systematic uncertainties in our $cD$ potential and $c\bar c$ potential.
One is in the short range region cased by the discretized Laplacian, and the second is the discretized lattice actions and the gauge fixing.
The last is from the reflection from the boundaries. 
The first two errors may be reduced by calculating with finer lattice and with improved actions and gauge fixing procedure.
Lattice setups with larger spatial extent is desirable for the third.

The quark-diquark potential and diquark mass in this study are obtained from lattice setup with $m_\pi\simeq700$ MeV.
 However, this is not the value at the physical point where the pion mass is $m_\pi\simeq140$ MeV. 
 In order to construct a diquark model which outputs observables that can be compared with experimental results, it is necessary to calculate the quark-diquark potential and the diquark mass at the physical point. 
 This is achieved by extrapolating to the physical point. 
 For example, we may calculate using available PACS-CS configurations corresponding to $m_\pi\simeq540$ MeV, $m_\pi\simeq410$ MeV, etc. and extrapolate the obtained results to the physical point.
  Physical observables such as levels of Baryon and exotic candidates calculated using the diquark model are comparable to experimental results.
  
Moreover, it is presumed that pseudo-scalar, vector and axial-vector diquarks may play crucial roles in a general context of hadron physics.
Straightforward extension of our method to the above mentioned diquarks seems to be an interesting step to pursue.
\section{\label{kannsha}Acknowledgement}
The author thanks N.~Ishii, A.~Hosaka, M.~Koma, A.~Nakamura, Tm.~Doi and K.~Sasaki   for fruitful
discussions.
The  Lattice  QCD  calculation  has  been carried  out  by  using  the
supercomputer OCTOPUS at Cyber Media  Center of Osaka University under
the  support  of   Research  Center  for  Nuclear   Physics  of  Osaka
University.
We also thank PACS-CS Collaboration and ILDG/JLDG for providing us with the
2+1          flavor          QCD         gauge          configurations
~\cite{pacs_config,beckett,ILDG,JLDG}.  The lattice QCD code is partly based on Bridge++~\cite{Bridge}.
This  research is  supported by MEXT  as  ``Program  for  Promoting Researches  on  the  Supercomputer
Fugaku''  (Simulation  for basic  science:  from  fundamental laws  of
particles to creation  of nuclei) and JICFuS and 
by JSPS KAKENHI Grant Number JP21K03535.

\appendix
\section{\label{appendix:m_q_spin_singlet} $m_q$ determination using the spin triplet sector}
To obtain the charm quark mass using the spin triplet sector, we start with replacing $\tilde V^{{\rm fit}}_{{\rm PS}}(r)$ in the Schr\"odinger equation \Eq{eq:Schrodinger_eq_radial} by $\widetilde{V}^{{\rm fit}}_{V}(r)=\widetilde{V}^{\rm fit}_{0}(r)+\frac{1}{4}\widetilde{V}^{\rm fit}_{\sigma}(r)$.

\begin{figure}[h]
  \centering
\includegraphics[width=0.48\textwidth]{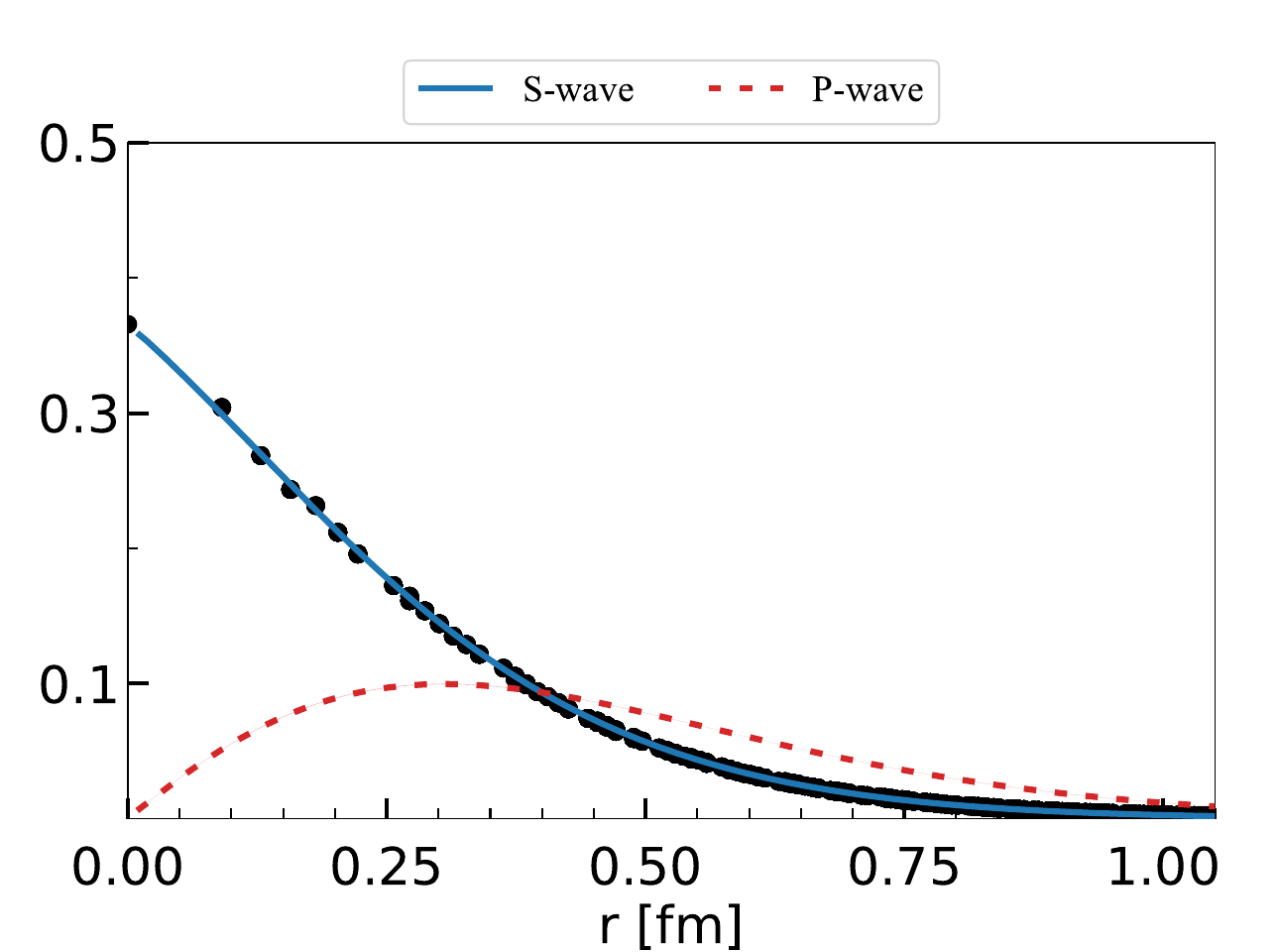}%
\caption{\label{ccbar_V_num}Numerical  solutions for S-wave (S) and P-wave (P).
	Black circles denote the NBS wave function data.}
\end{figure}

We use the pre-potential fit to Series \rmII in \Sect{Meson_Num}.
Then the spin-orbit averaged P-wave excitation pre-energy is obtained by incorporating the centrifugal potential as was demonstrated in \Eq{eq:pre-schroedinger} for the spin singlet sector.
\Fig{ccbar_V_num} compares the V channel numerical solutions to the NBS wave function where we find a good agreement.

Next, we replace $M_{{\rm T}} - M_{{\rm PS}}$ by $2M_{{\rm AV}}-M_{{\rm S}}-M_{{\rm V}}$ in \Eq{charm_mass} where we have neglected the small contribution from the tensor interaction.
Then, by substituting the P-wave pre-excitation energy into \Eq{charm_mass}, we get charm quark mass $1.604(76)$ GeV, which is consistent with that from the singlet sector within the statistical error.
Note that for better precision, the energy level of $2^{++}$ state ($\chi_{c2}$ meson) needs to be taken into account to offset the effect of the tensor interaction.

\section{Cubic group transformations\label{App:cubic}}
The cubic group has 48 elements, each being products of three $\frac{\pi}{2}$-rotations about $x$, $y$ and $z$ axes and the parity transformation.
The elements are representable by the parity-inversion and the order 4 sub-group matrices $C_x$, $C_y$ and $C_z$ given by
\begin{widetext}
\begin{equation}
\begin{array}{lll}
C_x = \pmat{
&\ 1& &\ 0& &\ 0& \\ 
&\ 0& &\ 0& &-1&\\
&\ 0& &\ 1& &\ 0&
	},
C_y = \pmat{
&\ 0& &\ 0& &\ 1& \\ 
&\ 0& &\ 1& &\ 0&\\
&-1& &\ 0& &\ 0&
	},
C_z= \pmat{
&\ 0& &-1& &\ 0& \\ 
&\ 1& &\ 0& &\ 0&\\
&\ 0& &\ 0& &\ 1&
	}
\end{array}
\label{def:rank_4_elements}
.
\end{equation}
\end{widetext}
\Table{table:cubic_list} summarizes the 24 proper rotations.
Combining these and the parity-inversion, we get 48 elements of $O_h$.

\begin{table}[h]
	\caption{24 matrices of proper rotation and the order of each element.}
	\begin{center}
		\begin{tabular}{c |  c } 
		Order  	&	Matrices	\\ \hline &\\
		1	&	$C_0\equiv \rm{unity}$ \\ &\\
		4	&	$\begin{array}{cccc}
							C_x,&C_y,&C_z&
							\\
							C_x^{-1},&C_y^{-1},&C_z^{-1},&
						\end{array}$ \\ &\\
		3	&	$\begin{array}{ccccc}
							C_xC_y,&C_zC_y,&C_xC_z,&C_yC_x&
							\\
							(C_xC_y)^2,&(C_yC_x)^2,&(C_zC_y)^2,&(C_xC_z)^2&
						\end{array}$ \\ &\\
		2	&	$C_x^2,C_y^2,C_z^2$ \\ &\\
		2	&	$\begin{array}{cccc}
							C_xC_yC_z,&C_yC_zC_x,&C_zC_xC_y
							\\
							C_x^{-1}C_y^{-1}C_z,&C_y^{-1}C_z^{-1}C_x,&C_z^{-1}C_x^{-1}C_y
						\end{array}$
		\end{tabular}
	\end{center}
	\label{table:cubic_list}
\end{table}
\bibliographystyle{apsrev4-2}
\bibliography{biblist.bib}
\end{document}